\documentclass[aps,prd,floatfix,notitlepage,noshowpacs,preprintnumbers,nofootinbib,superscriptaddress,twocolumn]{revtex4-2}
\usepackage[utf8]{inputenc}
\usepackage{amsmath}
\usepackage{amssymb}
\usepackage[T1]{fontenc}
\usepackage{bbold}

\usepackage{enumerate}
\usepackage{amsmath}
\usepackage{tikz}
\usepackage[compat=1.1.0]{tikz-feynman}
\usepackage{xcolor,colortbl}
\usepackage{feynmf}
\usepackage{slashed}
\usepackage{braket}
\usepackage{url}
\usepackage{natbib}
\usepackage{graphicx}
\usepackage[pdftex,bookmarks,linktocpage,pdfpagelabels,plainpages=false,hyperfigures,linkcolor=blue,citecolor=blue]{hyperref} %for hypertext links
\hypersetup{colorlinks=true, urlcolor=teal}

\usepackage{mathrsfs,amssymb}  
\usepackage{cancel}
\usepackage[normalem]{ulem}

\definecolor{LightRed}{rgb}{1,0.7,0.7}

\begin{document}
\bibliographystyle{apsrev4-1}
\title{Testing Meson Portal Dark Sector Solutions to the MiniBooNE Anomaly at CCM}

\affiliation{Bartoszek~Engineering,~Aurora,~IL~60506,~USA}
\affiliation{Columbia~University,~New~York,~NY~10027,~USA}
\affiliation{University~of~Edinburgh,~Edinburgh,~United~Kingdom}
\affiliation{Embry$-$Riddle~Aeronautical~University,~Prescott,~AZ~86301,~USA }
\affiliation{University~of~Florida,~Gainesville,~FL~32611,~USA}
\affiliation{Los~Alamos~National~Laboratory,~Los~Alamos,~NM~87545,~USA}
\affiliation{Massachusetts~Institute~of~Technology,~Cambridge,~MA~02139,~USA}
\affiliation{Universidad~Nacional~Aut\'{o}noma~de~M\'{e}xico,~CDMX~04510,~M\'{e}xico}
\affiliation{University~of~New~Mexico,~Albuquerque,~NM~87131,~USA}
\affiliation{New~Mexico~State~University,~Las~Cruces,~NM~88003,~USA}
\affiliation{Texas~A$\&$M~University,~College~Station,~TX~77843,~USA}

\author{A.A.~Aguilar-Arevalo}
\affiliation{Universidad~Nacional~Aut\'{o}noma~de~M\'{e}xico,~CDMX~04510,~M\'{e}xico}
\author{S.~Biedron}
\affiliation{University~of~New~Mexico,~Albuquerque,~NM~87131,~USA}
\author{J.~Boissevain}
\affiliation{Bartoszek~Engineering,~Aurora,~IL~60506,~USA}
\author{M.~Borrego}
\affiliation{Los~Alamos~National~Laboratory,~Los~Alamos,~NM~87545,~USA}
\author{L.~Bugel}
\affiliation{Massachusetts~Institute~of~Technology,~Cambridge,~MA~02139,~USA}
\author{M.~Chavez-Estrada}
\affiliation{Universidad~Nacional~Aut\'{o}noma~de~M\'{e}xico,~CDMX~04510,~M\'{e}xico}
\author{J.M.~Conrad}
\affiliation{Massachusetts~Institute~of~Technology,~Cambridge,~MA~02139,~USA}
\author{R.L.~Cooper}
\affiliation{Los~Alamos~National~Laboratory,~Los~Alamos,~NM~87545,~USA}
\affiliation{New~Mexico~State~University,~Las~Cruces,~NM~88003,~USA}
\author{A.~Diaz}
\affiliation{Massachusetts~Institute~of~Technology,~Cambridge,~MA~02139,~USA}
\author{J.R.~Distel}
\affiliation{Los~Alamos~National~Laboratory,~Los~Alamos,~NM~87545,~USA}
\author{J.C.~D’Olivo}
\affiliation{Universidad~Nacional~Aut\'{o}noma~de~M\'{e}xico,~CDMX~04510,~M\'{e}xico}
\author{E.~Dunton}
\affiliation{Columbia~University,~New~York,~NY~10027,~USA}
\author{B.~Dutta}
\affiliation{Texas~A$\&$M~University,~College~Station,~TX~77843,~USA}
\author{D.~Fields}
\affiliation{University~of~New~Mexico,~Albuquerque,~NM~87131,~USA}
\author{J.R.~Gochanour}
\affiliation{Los~Alamos~National~Laboratory,~Los~Alamos,~NM~87545,~USA}
\author{M.~Gold}
\affiliation{University~of~New~Mexico,~Albuquerque,~NM~87131,~USA}
\author{E.~Guardincerri}
\affiliation{Los~Alamos~National~Laboratory,~Los~Alamos,~NM~87545,~USA}
\author{E.C.~Huang}
\affiliation{Los~Alamos~National~Laboratory,~Los~Alamos,~NM~87545,~USA}
\author{N.~Kamp}
\affiliation{Massachusetts~Institute~of~Technology,~Cambridge,~MA~02139,~USA}
\author{D.~Kim}
\affiliation{Texas~A$\&$M~University,~College~Station,~TX~77843,~USA}
\author{K.~Knickerbocker}
\affiliation{Los~Alamos~National~Laboratory,~Los~Alamos,~NM~87545,~USA}
\author{W.C.~Louis}
\affiliation{Los~Alamos~National~Laboratory,~Los~Alamos,~NM~87545,~USA}
\author{J.T.M.~Lyles}
\affiliation{Los~Alamos~National~Laboratory,~Los~Alamos,~NM~87545,~USA}
\author{R.~Mahapatra}
\affiliation{Texas~A$\&$M~University,~College~Station,~TX~77843,~USA}
\author{S.~Maludze}
\affiliation{Texas~A$\&$M~University,~College~Station,~TX~77843,~USA}
\author{J.~Mirabal}
\affiliation{Los~Alamos~National~Laboratory,~Los~Alamos,~NM~87545,~USA}
\author{D.~Newmark}
\affiliation{Massachusetts~Institute~of~Technology,~Cambridge,~MA~02139,~USA}
\author{P.~deNiverville}
\affiliation{Los~Alamos~National~Laboratory,~Los~Alamos,~NM~87545,~USA}
\author{V.~Pandey}
\affiliation{University~of~Florida,~Gainesville,~FL~32611,~USA}
\author{D.~Poulson}
\affiliation{Los~Alamos~National~Laboratory,~Los~Alamos,~NM~87545,~USA}
\author{H.~Ray}
\affiliation{University~of~Florida,~Gainesville,~FL~32611,~USA}
\author{E.~Renner}
\affiliation{Los~Alamos~National~Laboratory,~Los~Alamos,~NM~87545,~USA}
\author{T.J.~Schaub}
\affiliation{University~of~New~Mexico,~Albuquerque,~NM~87131,~USA}
\author{A.~Schneider}
\affiliation{Massachusetts~Institute~of~Technology,~Cambridge,~MA~02139,~USA}
\author{M.H.~Shaevitz}
\affiliation{Columbia~University,~New~York,~NY~10027,~USA}
\author{D.~Smith}
\affiliation{Embry$-$Riddle~Aeronautical~University,~Prescott,~AZ~86301,~USA }
\author{W.~Sondheim}
\affiliation{Los~Alamos~National~Laboratory,~Los~Alamos,~NM~87545,~USA}
\author{A.M.~Szelc}
\affiliation{University~of~Edinburgh,~Edinburgh,~United~Kingdom}
\author{C.~Taylor}
\affiliation{Los~Alamos~National~Laboratory,~Los~Alamos,~NM~87545,~USA}
\author{A.~Thompson}
\affiliation{Northwestern~University,~Evanston,~IL~60208,~USA}
\author{W.H.~Thompson}
\affiliation{Los~Alamos~National~Laboratory,~Los~Alamos,~NM~87545,~USA}
\author{M.~Tripathi}
\affiliation{University~of~Florida,~Gainesville,~FL~32611,~USA}
\author{R.T.~Thornton}
\affiliation{Los~Alamos~National~Laboratory,~Los~Alamos,~NM~87545,~USA}
\author{R.~Van~Berg}
\affiliation{Bartoszek~Engineering,~Aurora,~IL~60506,~USA}
\author{R.G.~Van~de~Water}
\affiliation{Los~Alamos~National~Laboratory,~Los~Alamos,~NM~87545,~USA}

\begin{abstract}
    A solution to the MiniBooNE excess invoking rare three-body decays of the charged pions and kaons to new states in the MeV mass scale was recently proposed as a dark-sector explanation. This class of solution illuminates the fact that, while the charged pions were focused in the target-mode run, their decay products were isotropically suppressed in the beam-dump-mode run in which no excess was observed. This suggests a new physics solution correlated to the mesonic sector. We investigate an extended set of phenomenological models that can explain the MiniBooNE excess as a dark sector solution, utilizing long-lived particles that might be produced in the three-body decays of the charged mesons and the two-body anomalous decays of the neutral mesons. Over a broad set of interactions with the long-lived particles, we show that these scenarios can be compatible with constraints from LSND, KARMEN, and MicroBooNE, and evaluate the sensitivity of the ongoing and future data taken by the Coherent CAPTAIN Mills experiment (CCM) to a potential discovery in this parameter space. See addendum for updated predictions for future MicroBooNE sensitivity.
\end{abstract}

\preprint{LA-UR-23-29529}
\maketitle

\section{Introduction}
The excess of electron-like events observed by MiniBooNE~\cite{MiniBooNE:2008yuf,MiniBooNE:2018esg,MiniBooNE:2020pnu} at a level of $4.8\sigma$ has remained one of the stronger hints to the existence of new physics beyond the Standard Model (SM). The event data observed in the MiniBooNE detector is remarkable for its spectrum, with the excess appearing at forward scattering angles ($\cos\theta > 0.75$) and low energies ($E_{vis} < 500$ MeV), and for the asymmetry of excess events in the neutrino and anti-neutrino modes, while no excess was observed in the dump mode~\cite{MiniBooNEDM:2018cxm}, which had a reduced neutrino flux.

Neutrino-based new physics explanations have been popular solutions to the anomaly~\cite{Abdullahi:2023ejc,Sorel:2003hf,Karagiorgi:2009nb,Collin:2016aqd,Giunti:2011gz,Giunti:2011cp,Gariazzo:2017fdh,Boser:2019rta,Kopp:2011qd,Kopp:2013vaa,Dentler:2018sju,Abazajian:2012ys,Conrad:2012qt,Diaz:2019fwt,Asaadi:2017bhx,Karagiorgi:2012kw,Pas:2005rb,Doring:2018cob,Kostelecky:2003cr,Katori:2006mz,Diaz:2010ft,Diaz:2011ia,Gninenko:2009ks,Gninenko:2009yf,Bai:2015ztj,Moss:2017pur,Bertuzzo:2018itn,Ballett:2018ynz,Fischer:2019fbw,Moulai:2019gpi,Dentler:2019dhz,deGouvea:2019qre,Datta:2020auq,Dutta:2020scq,Abdallah:2020biq,Abdullahi:2020nyr,Liao:2016reh,Carena:2017qhd,Abdallah:2020vgg}. Since the neutrinos at MiniBooNE are produced primarily from charged meson decays and the decays of daughter muons of those charged mesons, neutrino-based solutions can accommodate the absence of any excess in the dump mode, in which the charged mesons are no longer focused by magnetic horns, unlike the neutrino and anti-neutrino modes. Essentially, the neutrino-based explanations work well because a key feature of the excess seems to be correlated to the focusing or suppression of charged mesons. Further, since the energy and angular distributions of the excess are already characteristic of neutrino-like signals and backgrounds, new physics appearing in the neutrino sector may naturally map onto the observed spectra.

This poses a challenge to dark sector interpretations of the excess (e.g., using $\pi^0$ or dark bremsstrahlung production channels~\cite{MiniBooNEDM:2018cxm,Jordan:2018qiy}), which have been more constrained and less complete in their explanation of the excess with respect to their counterparts in neutrino BSM physics thus far. Recently, a generic set of solutions to the excess was proposed in ref.~\cite{Dutta:2021cip} using a framework of rare three-body decays of the charged mesons -- decays which may not be strongly suppressed in their phase space and can be powerful probes of BSM physics~\cite{Barger:2011mt,Carlson:2012pc,Laha:2013xua,Bakhti:2017jhm,Altmannshofer:2019yji,Krnjaic:2019rsv}. One subset of models considered consists of a long-lived dark sector boson (not necessarily the cosmological DM, but at least long-lived on the scale of the BNB-to-MiniBooNE beamline of $\sim500$m) that can survive and scatter in the MiniBooNE detector via a photoconversion process, leaving a single photon in the final state. Since the Cherenkov detector does not distinguish between photons and electrons, this scattering process can be a viable contributor to the excess, provided that the appropriate phenomenological model can be found to be safe from existing constraints.

In this work, we will constrain the space of operators in an effective field theory (EFT) that leads to rare three-body decays of the charged mesons and photoconversion scattering of long-lived mediators at Coherent CAPTAIN Mills (CCM), utilizing the close proximity to the Lujan beam target as a source of stopped-pion decays. Using the CCM120 engineering run, we set conservative limits on the parameter space, and forecast sensitivity for the ongoing 3-year run with the upgraded CCM200 detector. A similar analysis for electromagnetic signal region of interest (ROI) that was performed for axion-like particles in ref.~\cite{CCM:2021jmk} is carried out here. In surveying the greater landscape of dark sector models that can explain the MiniBooNE anomaly via rare meson decays, we also take into account the analyses and observations from other stopped-pion experiments like LSND and KARMEN. Important findings about the parameter space that we consider here can be distilled from the existing data at MicroBooNE, and other forthcoming short-baseline experiments like SBND have a discovery potential here as we will show. In fact, the joint analysis of close-proximity stopped-pion experiments and those at the short-baseline neutrino program with magnetic horn-focused charged meson fluxes will have total experimental coverage over the parameter space explaining the anomaly.

This paper is organized as follows. In \S~\ref{sec:models} we introduce the operator EFT extension to the SM that we wish to consider and connect it to a phenomenological model of pion decays and photoconversion scattering in \S~\ref{sec:pheno}. In \S~\ref{sec:analysis} the analysis procedures for both MiniBooNE target and dump mode runs is discussed, in addition to our analysis of LSND, KARMEN, and MicroBooNE data interpreted as constraints on the models in question. In \S~\ref{sec:ccm} we show the analysis procedure for the CCM120 engineering run data and construct forecasts for the ongoing CCM200 data-taking run. In \S~\ref{sec:results} the resulting fits and constraints are shown for several benchmark models that utilize the operators we have considered in \S~\ref{sec:models}, and finally in \S~\ref{sec:conclude} we conclude.

\section{Models}
\label{sec:models}
We study a set of effective operators which permit, at a purely phenomenological level, the production of a long-lived particle (LLP) bosonic state from the three-body decay of the charged mesons, and subsequent photoconversion of said meson via a massive mediator; schematically,
\begin{equation}
\underbrace{\begin{tikzpicture}
        \begin{feynman}
            \vertex (a) {\(\pi^\pm\)};
            \vertex [right=1.2cm of a, blob] (b) {\(\, \, \, \, \, \)};
            \vertex [right=1.2cm of b] (f2) {\(X\)};
            %\vertex [right=1.0cm of f2] (p) {\(+\)};
            \vertex [below left=0.3cm of b] (xl);
            \vertex [above right=1.2cm of b] (f1) {\(\nu\)};
            \vertex [below right=1.2cm of b] (f3) {\(\ell\)};
            
            \diagram* {
                (a) -- [scalar] (b) -- [fermion] (f1),
                (b) -- [anti fermion] (f3),
                (b) -- (f2)
            };
        \end{feynman}
    \end{tikzpicture}}_\textrm{Beam Target Production}
    \begin{tikzpicture}
           \begin{feynman}
                    \vertex (o3) {\(\)};
                    \vertex [above=1.0cm of o3] (p) {\(+\)};
           \end{feynman}
    \end{tikzpicture}
    \underbrace{\begin{tikzpicture}
    \begin{feynman}
         \vertex (o1);
         \vertex [above left=0.7cm of o1] (f1) {\(X\)};
         \vertex [above right=0.7cm of o1] (i1){\(\gamma\)};
         \vertex [below=0.7cm of o1] (o2);
         \vertex [right=0.7cm of o2] (f2) {\(N\)};
         \vertex [left=0.7cm of o2] (i2) {\(N\)};

         \diagram* {
           (i1) -- [boson] (o1) -- (f1),
           (o1) -- [edge label={\(Y\)}] (o2),
           (i2) -- [fermion] (o2),
           (o2) -- [ fermion] (f2),
         };
        \end{feynman}
       \end{tikzpicture}}_\textrm{Detection}
      \nonumber
\end{equation}
This simple setup was shown to explain the MiniBooNE excess in ref.~\cite{Dutta:2021cip}, making use of two prominant features; (I) the coupling of a boson $X$ to the charged pion decays ensures the $X$ flux is correlated to the relative size of the excess in the target and off-target modes through the focusing of charged pions via the magnetic horns, and (II) the mass of the mediator $Y$ gives a dial to tune the angular spectrum of the outgoing $\gamma$'s Cherenkov ring, which is characteristically off-forward.

We will investigate a broad set of operators that allow for such a phenomenology in order to estimate the relative sizes of the parameter space allowed by existing constraints that also can accomodate the MiniBooNE excess. In \S.~\ref{sec:modelA} we consider a generic EFT for the two bosons $X$ and $Y$ below the QCD scale, while in \S.~\ref{sec:modelB} we consider a hadrophillic scenario with a single new boson and connect the EFT to specific quark couplings above the QCD scale.

\subsection{One Long-lived Boson and a Secondary Massive Mediator}\label{sec:modelA}
We study two BSM scenarios that each could explain the MiniBooNE excess and are testable at stopped pion and other beam dump facilities. In the first scenario, we extend the low energy SM EFT with two massive bosons, one of them long-lived and being produced via the three-body decay of the charged mesons and the other generally being heavier and facilitating photoconversion $X N \to \gamma N$. These decay and scattering mechanisms can arise from a multitude of operators. For the decays, scalars ($\phi$), pseudoscalars ($a$), or vectors ($V$) coupled to the electrons or muons through $g_\ell^s \phi \bar{\ell} \ell$, $-i g_\ell^p a \bar{\ell} \gamma^5 \ell$, and $g_\ell^V V_\mu \bar{\ell}\gamma^\mu \ell$ terms allow for $\pi^\pm \to X \ell \nu$ where $X = \phi, a, V$ is radiated off the charged lepton leg. Alternatively, effective couplings to the charged pions through operators like $g^s_\pi \phi \pi^+ \pi^-$ or $g^v_\pi V_\mu \pi^+ (\partial^\mu \pi^-)$ allow for radiative decays from the pion current (and potentially other contact and pion structure-dependent interactions, discussed in the next section).

On the detection side, the long-lived $\phi, a$, or $V$ mediators can induce single-photon final states through 
the dimension-5 couplings $\frac{\lambda^s}{4} \phi H_{\mu\nu} F^{\mu\nu}$ and $\frac{\lambda^p}{4} a H_{\mu\nu}\Tilde{F}^{\mu\nu}$, where we define $H_{\mu\nu} \equiv \partial_\mu V_\nu - \partial_\nu V_\mu$. In these cases, either a vector, scalar, or pseudoscalar can serve as the long-lived $X$ and another as the scattering mediator $Y$. Such operators can arise easily in concrete, UV-complete models. For example, they can fit within the framework of a $U(1)$ extension to the SM with extra fermions that permit a loop-induced coupling between a (pseduo)scalar, the $U(1)$ gauge boson, and the SM photon (see e.g. $U(1)_{T3R}$~\cite{Dutta:2022qvn} models or dark photon / axion portals~\cite{Kaneta:2016wvf}).
%with operators shown in Table~\ref{tab:ops},
%\begin{table}[h]
%    Decay Operators\\
%    \begin{tabular}{l|c c c}
%    \hline
%        $\mathcal{O}_\ell^{s,p,v}$ &  $g_\ell^s \phi \bar{\ell} \ell$ & $- i g_\ell^p a \bar{\ell} \gamma^5 \ell$ & $g_\ell^V V_\mu \bar{\ell}\gamma^\mu \ell$ \\
%        $\mathcal{O}_\pi^{s,p,v}$ & $g^s_\pi \phi \pi^+ \pi^-$ & $g^p_\pi a \pi^+ \pi^-$ & $g^v_\pi V_\mu \pi^+ (\partial^\mu \pi^-)$
%    \end{tabular}\\
%    Scattering Operators\\
%    \begin{tabular}{l|c c c}
%    \hline
%        $\mathcal{O}^{s,p}_5$ & $\frac{\lambda^s}{4} \phi H_{\mu\nu} F^{\mu\nu}$ & $\frac{\lambda^p}{4} a H_{\mu\nu}\Tilde{F}^{\mu\nu}$ & \\
%        $\mathcal{O}^{s,p,v}_n$ & $\phi \overline{N} \mathbf{Y}^s_n N$ & $i a \overline{N} \gamma^5 \mathbf{Y}^p_n N$ & $V_\mu \overline{N} \gamma^\mu \mathbf{Y}^v_n N$
%    \end{tabular}
%    \caption{Operators that allow for the three-body decay of charged mesons producing long-lived scalar ($\phi$), pseudoscalar ($a$), and vector ($V$) bosons and their subsequent photoconversion to a visible $\gamma$.
%    \label{tab:ops}
%\end{table}
%where $N = (p, n)$ is the nucleon doublet and $\mathbf{Y}^i_n = (y_n^{0,i} + y_n^{1,i} \tau_3)$ is the nucleon coupling matrix for $i=s,p,v$. We also define $H_{\mu\nu} \equiv \partial_\mu V_\nu - \partial_\nu V_\mu$.

\subsection{A Single Long-lived Boson Coupling to Quarks}\label{sec:modelB}
For this scenario, we consider a hadrophillic model that only couples to first generation quarks. Let's start with a massive vector boson, where above the QCD phase transition, its interactions with quarks is described by the Lagrangian
\begin{align}
    \mathcal{L} &\supset  \sum_{q=u,d} g_q V_\mu \overline{q} \gamma^\mu q + \textrm{h.c.}
\end{align}
We could interpret this as an extra $U(1)$ gauging the quarks with some dark charge, for example. Taking this below the QCD scale in the chiral perturbation theory ($\chi$PT), we have an operator like~\cite{Berger:2019aox}
\begin{equation}
\label{eq:chiPT-vector-pion}
    \mathcal{L}^{\chi PT} \supset ig_{\pi^\pm} V_\mu \pi^+ (\partial^\mu \pi^-)
\end{equation}
Additionally, the chiral anomaly will lead to the anomalous decay of the $\pi^0$ to $\gamma V$ through the dimension-5 operator
\begin{equation}
   \mathcal{L}^{\chi PT} \supset g_{\pi^0} \frac{e}{16\pi f_\pi}\pi^0 F_{\mu\nu}\Tilde{H}^{\mu\nu}
\end{equation}
We have defined $H_{\alpha\beta} = \partial_\alpha V_\beta - \partial_\beta V_\alpha$, and $\Tilde{H}^{\mu\nu} = \epsilon^{\mu\nu\alpha\beta}H_{\alpha\beta}$. The dimension-5 interaction $\sim \pi^0 F_{\mu\nu}\Tilde{H}^{\mu\nu}$ permits $X N \to \gamma N$ via the pion-nucleon interaction $i g_{\pi N} \pi^0 \bar{\psi} \gamma^5 \psi$. This is an interaction of strong coupling; $g_{\pi N} \simeq 13$~\cite{deSwart:1997ep,PhysRevC.95.064001}.\footnote{One might also consider nuclear couplings which permit $V N \to V N$ scattering, leaving a nuclear recoil signature, but these processes would be $\mathcal{O}(g_q^4)$ suppressed.}

\section{Phenomenology}
\label{sec:pheno}
\subsection{Three-body Decays of Charged Mesons}
\begin{figure}
        \begin{tikzpicture}
        \begin{feynman}
            \vertex (a) {\(\pi^\pm\)};
            \vertex [right=1.4cm of a] (b);
            \vertex [above right=0.5cm of b] (fm);
            \vertex [above right=1.4cm of b] (f1) {\(\ell\)};
            \vertex [below right=1.0cm of b] (c);
            \vertex [below left=0.1cm of c] (xl) {IB1};
            \vertex [above=1.0cm of fm] (f2) {\(X=\phi, a, V\)};
            \vertex [below right=0.4cm of c] (f3) {\(\nu\)};
            \diagram* {
            (a) -- [scalar] (b) -- (f1),
            (b) -- (f3),
            (fm) --  (f2),
            };
        \end{feynman}
    \end{tikzpicture}\\
    \begin{tikzpicture}
        \begin{feynman}[small]
            \vertex (a) {\(\pi^\pm\)};
            \vertex [right=1.0cm of a] (x);
            \vertex [below=0.05cm of x] (xl) {IB2};
            \vertex [above=0.8cm of x] (f2) {\(X=\phi, a, V\)};
            \vertex [right=1.6cm of a] (b);
            \vertex [above right=1.4cm of b] (f1) {\(\ell\)};
            \vertex [below right=1.4cm of b] (f3) {\(\nu\)};
            \diagram* {
            (a) -- [scalar] (b) -- (f1),
            (b) -- (f3),
            (x) --  (f2),
            };
        \end{feynman}
    \end{tikzpicture}\\
    \begin{tikzpicture}
        \begin{feynman}
            \vertex (a) {\(\pi^\pm\)};
            \vertex [right=1.6cm of a] (b);
            \vertex [above=1.1cm of b] (f2) {\(X=\phi, a, V\)};
            \vertex [below left=0.7cm of b] (xl) {IB3, SD};
            \vertex [above right=1.4cm of b] (f1) {\(\ell\)};
            \vertex [below right=1.4cm of b] (f3) {\(\nu\)};
            
            \diagram* {
                (a) -- [scalar] (b) -- (f1),
                (b) -- (f3),
                (b) -- (f2)
            };
        \end{feynman}
    \end{tikzpicture}
    \caption{3-body charged meson decay $\pi \to \ell \nu X$ for a bosonic Lorentz representation $X=\phi,a,V$.}
    \label{fig:3body}
\end{figure}
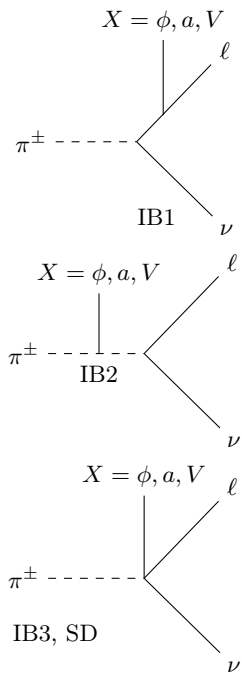

The three-body radiative decay of the $X$ boson off the pion current (Fig.~\ref{fig:3body}) can be modeled off the work on radiative meson decays in the SM, like $\pi \to \gamma e \nu$ (see for example refs.~\cite{Donoghue.PhysRevD.40.2378,BRYMAN1982151}). For the process $\pi(p) \to X(k) e(\ell) \nu(q)$, there are several types of so-called ``internal bremsstrahlung'' (IB) interactions, adopting the nomenclature of the aforementioned reference; IB1, radiating $X$ off the lepton leg; IB2, radiating off the charged meson current; and IB3, or a 4-point contact interaction radiating off the pion-lepton vertex. Additionally, we may also consider structure-dependent (SD) terms originating from mixing between the new boson $X$ and the vector mesons, but as these branching ratios are strongly suppressed, we will set them aside in this discussion.

Our discussion now focuses on a massive vector boson $V$ taking part in these radiative decays, but we will return to the case of $X = $ scalar, pseudoscalar later.

The matrix elements for each process can be described by factorizing the leptonic and hadronic parts of the current;
\begin{align}
    \mathcal{M} &= i \frac{G_F}{\sqrt{2}} \varepsilon^\mu \big[ \bar{u}_\ell \gamma^\rho (1-\gamma^5)v_\nu \big] T_{\mu\rho}
\end{align}
where the hadronic tensor $T_{\mu\rho}$ can be expressed in terms of the amplitude
\begin{equation}
    T_{\mu\rho} = i \int d^4 x e^{i k x} \braket{0 | T[j^V_\mu(x) j^+_\rho(0)] |\pi^+ (p)}
\end{equation}
for currents $j^V_\mu = \sum_q g_q \bar{q}\gamma_\mu q$ and $j^+_\rho = \bar{d} \gamma_\rho (1-\gamma^5) u$. The IB2 term shown in Fig.~\ref{fig:3body}, middle, is given by
\begin{equation}
        T_{\mu\rho}^\textrm{IB2} = \dfrac{i (2p-k)_\mu (p-k)_\rho }{2p\cdot k - m_V^2} f_\pi
\label{eq:IB2}
\end{equation}
which may come directly from the action in Eq.~\ref{eq:chiPT-vector-pion}, $i g_{\pi^\pm} V_\mu \pi^+ (\partial^\mu \pi^-)$, while additional contact and structure-dependent terms in Fig.~\ref{fig:3body}, bottom, may come from less trivial interactions. For example, a simple contact term could manifest from making the gauge covariant replacement $\partial_\mu \to \partial_\mu - i g_{\pi^\pm} V_\mu$ in the pion-lepton Fermi interaction $(\partial_\mu \pi^+) \bar{\ell} \gamma^\mu (1 - \gamma^5) \nu$.

However, we can only speculate about the gauge nature of our massive vector, and so to proceed naively we decompose the hadronic tensor $T_{\mu\rho}$ in a gauge covariant way. We follow the approach given in ref.~\cite{Khodjamirian:2001ga}, expressing $T_{\mu\rho}$ in terms of the momenta $k_\mu$ and $L_\mu \equiv q_\mu + \ell_\mu$;
\begin{align}
\label{eq:tmunu}
    T_{\mu\rho} &= \Tilde{a}_0 g_{\mu\rho} + \Tilde{b}_0 L_\mu k_\rho + \Tilde{b}_1 L_\rho k_\mu \nonumber \\
    &+ \Tilde{b}_2 L_\mu L_\rho + \Tilde{b}_3 k_\mu k_\rho + \epsilon_{\rho\mu\lambda\sigma} L^\lambda k^\sigma F_V
\end{align}
where $\Tilde{a}_0$, $\Tilde{b}_i$ ($i=0,1,2,3$), and $F_V$ are dimensionful invariant amplitudes. For massless photons, satisfying the Ward identity imposes a relationship between the coefficients~\cite{Khodjamirian:2001ga,Beneke:2018wjp,Beneke:2011nf, Bansal:2020xnb}.
%,namely that by taking a derivative of the amplitude and using the $T$-product identities, we get that
%\begin{equation}
%\label{eq:ward}
%    k^\mu T_{\mu\rho} = i p_\rho f_\pi = i (L + k)_\rho f_\pi .
%\end{equation}
However, for a massive vector boson $V^\mu$, no Ward identity needs to be satisfied in general, as a theory with massive vector bosons need not be gauge invariant.
%However, if one considers a St\"{u}ckelberg action whose vector field has a symmetry under $\delta V_\mu = \partial_\mu \Lambda(x)$ (and a corresponding symmetry for the associated scalar St\"{u}ckelberg field), then the Ward identity is restored. Proceeding in this more conservative manner, we assume Eq.~\ref{eq:ward} is a constraint.
%Applying $k^\mu T_{\mu\rho}$ explicitly to Eq.~\ref{eq:tmunu}, we have
%\begin{align}
%\Tilde{a}_0 + \Tilde{b}_0 (L\cdot k) + \Tilde{b}_3 m_V^2 &= i f_\pi \\
%\Tilde{b}_1 m_V^2 + \Tilde{b}_2 (L \cdot k) &= i f_\pi
%\end{align}
Noting that $L \cdot k = (p - k)\cdot k$, the IB2 term in Eq.~\ref{eq:IB2} is recovered if one takes $\Tilde{b}_2 = 2 \Tilde{b}_1 = i f_\pi / (2 L\cdot k + m_V^2)$. The remaining terms $\Tilde{a}_0, \Tilde{b}_0$, $\Tilde{b}_2$ and $F_V$ account for contact and structure dependent contributions. For example, a pure contact term would take the form
\begin{equation}
\label{eq:IB3}
    T_{\mu\rho}^\text{IB3} = i f_\pi \alpha g_{\mu\rho} + i \beta \frac{f_\pi}{(L\cdot k)} g_{\mu\rho}
\end{equation}
for dimensionless constants $\alpha, \beta$. In this analysis we will consider only two instances of the meson couplings; those of IB2 nature (Fig.~\ref{fig:3body}, middle, or Eq.~\ref{eq:IB2}) and those of a pure contact or IB3 nature (Fig.~\ref{fig:3body}, bottom or Eq.~\ref{eq:IB3}).

\subsection{Dark Boson Photoconversion}
\begin{figure}[h]
 \centering
  %%% Scattering
        \begin{tikzpicture}
              \begin{feynman}
         \vertex (o1);
         \vertex [above left=1.2cm of o1] (f1) {\(V\)};
         \vertex [above right=1.2cm of o1] (i1){\(\gamma\)} ;
         \vertex [below=1.2cm of o1] (o2);
         \vertex [right=1.2cm of o2] (f2) {\(N\)};
         \vertex [left=1.2cm of o2] (i2) {\(N\)};

         \diagram* {
           (i1) -- [boson] (o1) -- [boson] (f1),
           (o1) -- [scalar, edge label={\(\phi,a,\pi^0\)}] (o2),
           (i2) -- [fermion] (o2),
           (o2) -- [ fermion] (f2),
         };
        \end{feynman}
       \end{tikzpicture}
       \begin{tikzpicture}
              \begin{feynman}
         \vertex (o1);
         \vertex [above left=1.2cm of o1] (f1) {\(\phi,a\)};
         \vertex [above right=1.2cm of o1] (i1){\(\gamma\)} ;
         \vertex [below=1.2cm of o1] (o2);
         \vertex [right=1.2cm of o2] (f2) {\(N\)};
         \vertex [left=1.2cm of o2] (i2) {\(N\)};

         \diagram* {
           (i1) -- [boson] (o1) -- [scalar] (f1),
           (o1) -- [boson, edge label={\(V\)}] (o2),
           (i2) -- [fermion] (o2),
           (o2) -- [ fermion] (f2),
         };
        \end{feynman}
       \end{tikzpicture}
    \caption{Left: Vector photoconversion via a massive scalar or pseudoscalar mediator. Right: Scalar or pseudoscalar photoconversion via a massive vector mediator.}
    \label{fig:dark_primakoff}
\end{figure}
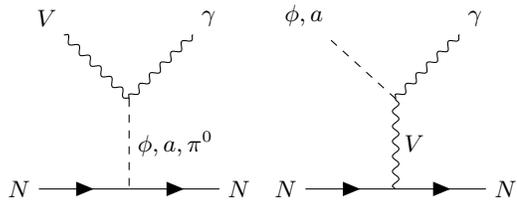

Long-lived bosonic states may have dimension-5 couplings between a secondary dark boson and the SM photon. This operator may arise e.g. at the 1-loop level from a theory connecting fermions charged under $U(1)_{em}$ to additional scalar and vector fields, or from the $\pi^0 \gamma X$ vertex in our single-mediator, hadrophillic scenario at the $\chi PT$ level. This opens a scattering channel similar to axion photoconversion or Primakoff scattering, except instead of the SM photon being in the $t$-channel, the secondary massive boson takes its place. This process $X N \to \gamma N$ (see Fig.~\ref{fig:dark_primakoff}) may be coherent if the mediator for the scattering couples to nucleons, provided that the momentum transfer scale $q \lesssim 1/R_n$ for a nuclear size $R_n$ and that the sum of the neutron and proton charges coupled to the mediator is not small, so that the total amplitude picks up a coherent enhancement $\propto (\sum_p Q_p + \sum_n Q_n)^2$. For simplicity we will assume that this coherent enhancement goes according to the proton number squared, $Z^2$, although depending on the baryonic couplings it could be larger or smaller (for instance, in the case of negative couplings). 

In the case of scalar photoconversion on a nucleus of mass $M$ via a heavy vector mediator $V$, we have
\begin{align}
\braket{|\mathcal{M}|^2}_\textrm{V}^{\phi\to\gamma} &= (g_n \lambda^s)^2 t \big[2 M^2 (m_\phi^2-2 s-t) +2 M^4 \nonumber \\
    &-2 m_\phi^2 (s+t)+m_\phi^4+2 s^2+2 s t+t^2\big] \nonumber \\
    & \times \frac{F_N^2(t)}{(8 (m_V^2-t)^2)}
\end{align}
where $s,t$ are the Mandelstam invariants for the center of momentum energy and momentum transfer, respectively. The nuclear form factor $F_N^2(t)$, for which we take the well-known Helm parameterization with normalization $F_N^2(0) = Z^2$. The same matrix element holds in the case of pseudoscalar photoconversion $a\to\gamma$. For vector photoconversion via a heavy scalar or pseudoscalar mediator, the spin-averaged matrix elements are
\begin{align}
    \braket{|\mathcal{M}|^2}_\textrm{S}^{V\to\gamma} &=\frac{3 (g_n^s \lambda^s)^2 \left(4 M^2-t\right)  \left(m_V^2-t\right)^2}{16 \left(m_\phi^2-t\right)^2} F_N^2(t)\\
    \braket{|\mathcal{M}|^2}_\textrm{P}^{V\to\gamma} &=\frac{(g_n^p \lambda^p)^2 \left(-t\right) \left(m_V^2-t\right)^2}{8 \left(m_a^2-t\right)^2} F_N^2(t)
\end{align}
For each case we check that the free matrix element is peaked in the momentum transfer, $t$, well within the coherent regime of momentum transfers where the Helm form factor is flat and unsuppressed. We find that the matrix element only starts to leave the coherent regime for large incoming energies $E_X \gtrsim 500$ MeV and for heavy mediator masses $m_Y \gtrsim 300$ MeV. Above this energy scale and above this heavy mediator limit, there may be a growing inelastic/incoherent component, but to avoid theoretical complications in this crossover regime between coherence and incoherence, we primarily consider the $m_Y \lesssim 300$ MeV regime.

\begin{figure}
    \centering
    \includegraphics[width=0.49\textwidth]{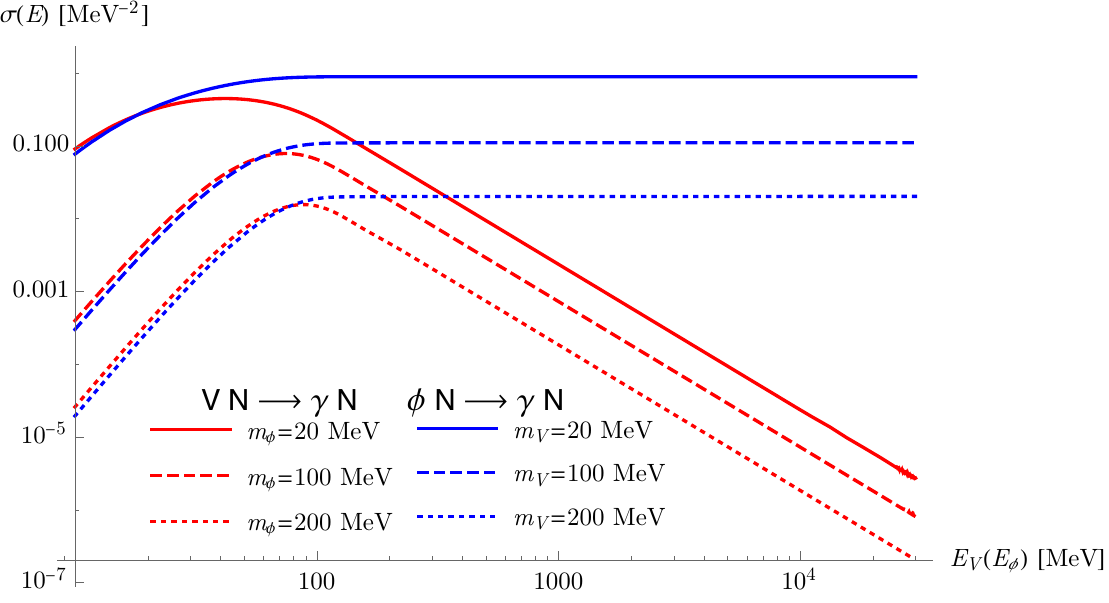}
    \caption{Cross sections for the photoconversion of a massive vector (scalar) mediated by a massive scalar (vector) through the dimension-5 vertex. We fix the mass of the incoming boson to 1 MeV in each case. The cross sections in the case of the photoconversion of a massive pseudoscalar are similar, varying only up to a constant factor.}
    \label{fig:xs}
\end{figure}

The cross sections associated with these matrix elements are shown in Fig.~\ref{fig:xs}. Notice that since in the case of an incoming vector photoconverting via a heavy scalar or pseudoscalar, $\braket{|\mathcal{M}|^2}_\textrm{S,P}^{V\to\gamma}$ has no $s$ dependence, and therefore the total cross section picks up its $s$ dependence only as $\sim 1/s^2$ from the Lorentz invariant phase space integration. The phenomenological impact of this difference between (pseudo)scalar-mediated and vector-mediated photoconversion is seen in Fig.~\ref{fig:xs} as either a decreasing or constant cross section as a function of the energy of the incoming boson, thereby impacting the fit at MiniBooNE in the high-energy / low-energy bins.

Finally, let us discuss the last possibility which arises from the effective dimension-5 coupling of a massive vector to the pion anomalous decay; $\mathcal{L} \supset \pi^0 F_{\mu\nu}\Tilde{H}^{\mu\nu}$. This effective interaction vertex typically appears in any $U(1)_X$ model where $V$ interacts with the SM quarks, resulting in a $\pi^0 - \gamma - V$ vertex at the pion level. Models involving this interaction vertex that satisfy the MiniBooNE excess are discussed in ref.~\cite{Dutta:2021cip}. This interaction permits $V N \to \gamma N$ via the pion-nucleon interaction $i g_{\pi N N} \pi^0 \bar{\psi} \gamma^5 \psi$ where the $\pi N N$ coupling is estimated around $g_{\pi N N} \simeq 13$~\cite{deSwart:1997ep,PhysRevC.95.064001}. 

Given that the neutral pion coupling $g_{\pi N N} \bar{N} \gamma^5 \tau_3 N \pi^0$ is opposite in sign for the proton and neutron, which have opposing isospin charges, $V N \to \gamma N$ scattering via coherent $\pi^0$ exchange is suppressed for most isotopes ($\propto (A-2Z)^2$). Instead, we consider single-nucleon scattering such that the process is incoherent and proportional to $A F_p^2(t)$, where $F_p(t)$ is the proton form factor, and we take $M = m_p, m_n$.

In the absence of a full nuclear model, we approximate this scattering on free nucleon initial and final states. Since we are primarily concerned with the electromagnetic component of the signal to explain the MiniBooNE anomaly, here we only take the final state photon as the visible energy and do not treat the physics of the nucleon final state. For the low energy excess, we find that most of the final state energy is carried by the photon. However, this nucleon final state may be relevant for studies in LArTPCs that can identify the tracks of ejected nucleons, so we leave this to a future study. Additionally, whether the final state nucleon is ejected or not, it may be likely that the nucleus is left in an excited state. The de-excitation photons released from nuclear transitions may also contribute to the signal, but with only $\mathcal{O}(1\,\text{MeV})$ of energy. This signal component may not be relevant for experiments with high energy thresholds (e.g. MiniBooNE, MicroBooNE) but could be relevant for liquid scintillators and especially the LSND anomaly. We leave this subject to a future work.

\section{Analysis of the MiniBooNE Data and Constraints}
\label{sec:analysis}
%In this section we discuss the fit procedure to derive the viable parameter space in the phenomenological models to explain the combined MiniBooNE target= and dump-mode data, making use of both angular and energy observables in the $\nu$, $\overline{\nu}$, and dump-mode datasets. We also discuss constraints that may come from the null observations of LSND and KARMEN data, and lastly, we discuss the analysis of the 6-week CCM120 engineering run data to constrain the same parameter space and show projections for the ongoing CCM200 data-taking campaign.

\subsection{MiniBooNE Target-mode and Beam-dump-mode Data}
We simulate the LLP flux and event spectra in the MiniBooNE detector by first modeling the flux of charged pions focused by the magnetic horns. This involves a detailed simulation of the Lorentz forces acting on the charged pions and their radiative transport through the horn system, discussed more in Appendix~\ref{app:horn}. Once the charged pion decays are modeled with the decay channels discussed in the previous section, the LLPs produced in these decays are propagated towards the detector and integrated over the its geometrical angular acceptance. This is done for forward and reverse horn currents, corresponding the neutrino and anti-neutrino mode data, respectively~\cite{MiniBooNE:2008yuf,MiniBooNE:2018esg,MiniBooNE:2020pnu}, and separately for the charged and neutral pion decays in the MiniBooNE beam dump without horn focusing~\cite{MiniBooNEDM:2018cxm}. Since the timing structure of the excess falls within 7 ns relative to the neutrino time-of-flight, we expect LLP masses above $\simeq 20$ MeV will begin to be less consistent with this timing structure due to their smaller boost factors. We do not employ a hard timing cut in this work, but instead limit our scope to masses below 35 MeV, above which this effect should become prominent. Their subsequent scattering via photonconversion processes $X N \to \gamma N$ give the distribution of $E_{vis}$ (for which we take equal to $E_\gamma$ for simplicity, although in principle a smearing matrix should be applied to more diligently model the detector resolution) and $\cos\theta_\gamma$ for the reconstructed Cherenkov rings. Given a set of couplings in the decay and scattering models, and the masses of the LLP and the scattering mediator, we then derive fits to the MiniBooNE data.

\begin{figure*}[ht]
    \centering
    \includegraphics[width=1.0\textwidth]{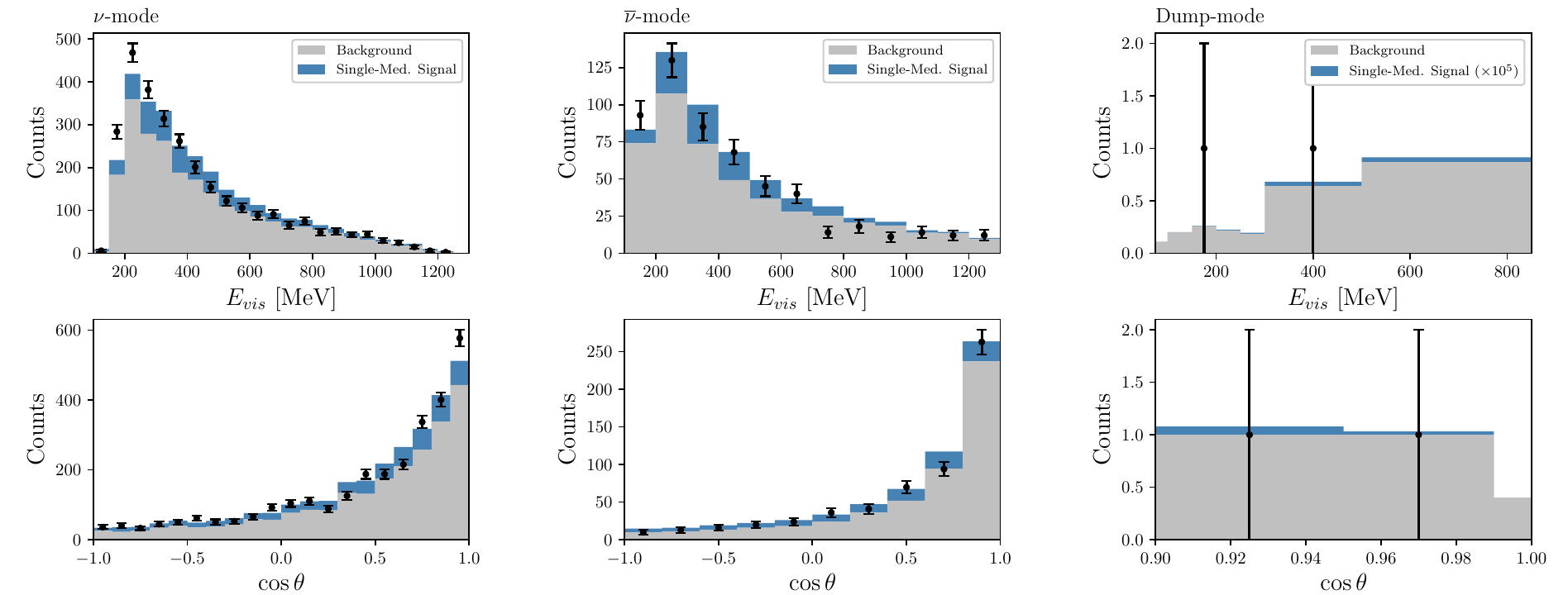}
    \caption{An example fit to the energy and cosine spectra in the MiniBooNE $\nu$-mode (left), $\overline{\nu}$-mode (center), and beam-dump-mode (right) with a long-lived vector $(V)$ produced from the IB2 decay of the charged pions and scattering into a single-photon final state via the SM $\pi^0$-nucleon interaction (see Eq.~\ref{eq:singlemedLag}). Here the signal in the beam-dump-mode is scaled up by a factor of $10^5$ for visualization.}
    \label{fig:mb_fit}
\end{figure*}

Example fits to the $\nu$-mode cosine and visible energy spectra are shown in Fig.~\ref{fig:mb_fit}. In the absence of full 2-dimensional data across $(E_{vis}, \cos\theta)$ and covariance matrices for both neutrino-mode and anti-neutrino-mode data, we compute a binned $\chi^2$ for both the visible energy data and cosine data for $N$ bins;
\begin{align}
    \chi^2_\nu &= \sum_{i=1}^{N} \frac{(d_i - s_i - b_i)^2}{\sigma_i^2}.
\end{align}
We then pick the more constraining of the two, either from the cosine or the visible energy data, to set the confidence levels.
A similar $\chi^2$ is constructed for the anti-neutrino-mode data and the beam-dump-mode data, and we combine all three data sets together in a joint $\chi^2$;
\begin{equation}
    \chi^2_\textrm{MB} = \chi^2_\nu + \chi^2_{\overline{\nu}} + \chi^2_\textrm{dump}
\end{equation}
In each model, the signal yield will be schematically proportional to branching ratio in the 3-body decay times scattering cross section, the yield will scale with the coupling product $g^2 \lambda^2 y^2$. For the operator combinations in model A, we generally fix the mass of the long-lived boson and allow the coupling product and the mass of the mediator in the photoconversion scattering to float in the fit.

It is important to note that the MiniBooNE excess is in time with the Booster Neutrino Beamline 52\,MHz beam timing structure \cite{MiniBooNE:2020pnu}, strongly suggesting that the source of the excess is relativistic.  This is to be expected from neutrinos or other light particle propagation (studied in this paper) from the target to the detector.

\subsection{MicroBooNE $1\gamma 0 p$ Data}
\begin{figure}
    \centering
    \includegraphics[width=0.5\textwidth]{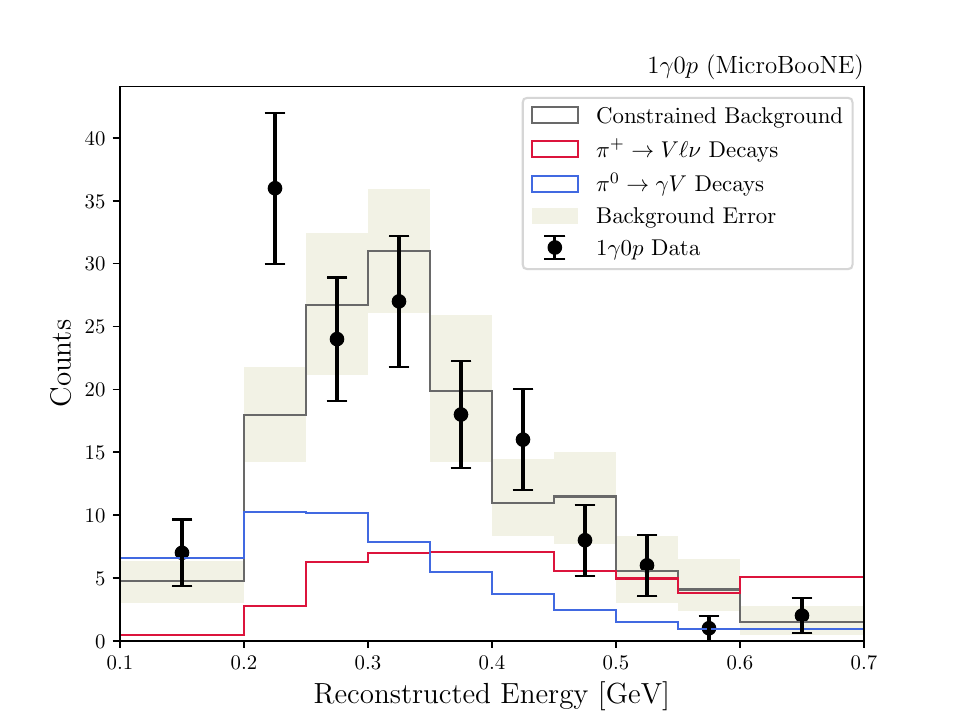}
    \caption{Example event spectra prediction from 2-body decays of $\pi^0 \to \gamma V$ (blue) and 3-body decays of $\pi^+ \to \ell \nu V$ (red) at MicroBooNE in the 1$\gamma$0p topology. Here the LLP mass $m_V$ is fixed to 5 MeV.}
    \label{fig:microboone_events}
\end{figure}

The MicroBooNE collaboration performed an analysis of $\Delta \to N \gamma$ resonant production utilizing several final state topologies, namely $1\gamma 0 p$ and $1\gamma 1 p$~\cite{MicroBooNE:2021zai}. We calculate the expected event rate at MicroBooNE, again employing the simulation procedure for the charged and neutral pions produced in the BNB target and focused through the horn system working in neutrino-mode polarization. This procedure follows exactly as in the previous section for the MiniBooNE analysis, as described in Appendix~\ref{app:horn}, except we now integrate the pion decay products over the solid angle spanned by the MicroBooNE detector's geometric cross section. We also apply the MicroBooNE-reported 5.29\% signal efficiency for this search. In Fig.~\ref{fig:microboone_events} we show example event spectra produced from 3-body charged meson decays as well as from 2-body $\pi^0$ decays using the VIB2 interaction model with couplings to the pion doublet.

One could also investigate the possibility of using the existing data of higher-energy beam dump experiments like CHARM (400 GeV) or MINER$\nu$A (120 GeV) to constrain the model parameter space. In the case of CHARM, we estimate $\mathcal{O}(10^{17})$ pions produced for $2\times 10^{18}$ collected POT~\cite{BERGSMA1985458}, and for a detector proximity of 480 meters, the expected flux of LLPs above the 5 GeV energy threshold should be comparatively smaller than those from LSND and KARMEN. Similarly, for MINER$\nu$A, one might examine the $\nu-e$ elastic scattering cross section measurements for events that would mimic the $1\gamma$ final state considered in our phenomenological models~\cite{MINERvA:2019hhc}. Although the beam energy at MINER$\nu$A is larger (NuMI beam) than the BNB flux at MicroBooNE, the detector tonnage at the latter is bigger given comparable collected POT and detector baselines. Hence we limit the scope of this analysis to deriving the more stringent constraints from the null results of the MicroBooNE data.

\subsection{Constraints from the LSND and KARMEN Null Results}
The parameter space associated with these scenarios get constrained by the LSND data. The LSND experiment used a 800-MeV proton beam. Three analyses, $e^- -\nu_e$ elastic scattering~\cite{LSND:2001akn}, charged current reactions of $\nu_e$ on $^{12}$C~\cite{LSND:2001fbw}, and neutrinos from the pion decay in flight~\cite{LSND:1997vqj} can be used to obtain constraints for the parameter space relevant to this solution to the MiniBooNE excess. The data from the elastic and inelastic analyses provide a constraint for the electromagnetic energy in the range 18-35 MeV while the decay-in-flight analysis provides a constraint for the energy range 60-200 MeV. 

A summary of the efficiencies and observed counts in each channel is given in Table~\ref{tab:lsnd}. To determine the constraints set by the null results of each channel on the parameter space of the decay and scattering models we consider, we adopt a single-bin $\chi^2$ as a crude test statistic. For example, using the DAR analysis, we look for a contour of constant $\Delta \chi^2 = s^2 / 1081$, where $s$ is the expected $X N \to \gamma N$ events (multiplied by a flat 37\% efficiency) in the energy range $[18, 35]$ MeV and 1081 is the number of observed events in the DAR region of interest.
\begin{table}[h]
    \centering
    \begin{tabular}{l|c|c|c|c}
         Analysis & $E_{vis}$ Range & $\cos\theta$ Range & Efficiency & Counts  \\
         \hline
         DAR & [$18,35$] MeV & $-1\leq\cos\theta \leq 1$ & 37\% & 1081 \\
         DIF & [$60,200$] MeV & $\cos\theta < 0.8$ & 10\% & 50
    \end{tabular}
    \caption{Kinematic regions of interest in the LSND decay-at-rest (DAR) and decay-in-flight (DIF) analyses, their signal efficiencies, and the target number of statistically significant signal counts.}
    \label{tab:lsnd}
\end{table}
Finally, it may be important to additionally consider inelastic responses of the nucleus in $X N \to \gamma N$ scattering. Although the scope of this work is limited to the null results of LSND, one might also attempt to explain the LSND excess via the inelastic scattering of $X N \to \gamma N^*, N^* \to N \gamma$, which, as mentioned in \S~\ref{sec:pheno}, would show up as gamma signal from nuclear de-excitation but will be left to a future work. 

Next, we can apply the KARMEN experiment's observations of the neutral current excitation process $^{12}$C$(\nu,\nu^\prime) \to ^{12}$C$(1^+,1;15.1$ MeV$\gamma)$ to place a constraint on photon final states arising from the photoconversion scattering in our phenomenological models~\cite{Karmen199815}. This data consists of $4.6 \times 10^{22}$ collected POT on the tungsten target at the ISIS~\cite{Reichenbacher:2005nc,Zeitnitz:1994kz}. The KARMEN detector was situated 17.5m from the target and totaled 56t of liquid scintillating hydrocarbon in a 3.5m$\times$4m$\times$4m geometry. To recast the NC analysis in ref.~\cite{Karmen199815} for our signal model, we will assume the same 12\% signal efficiency and 11.5\% energy resolution.

\section{Analysis of the CCM120 Data and CCM200 Projections}
\label{sec:ccm}

In 2019 a six week engineering beam run was performed with the CCM120 detector, named due to it having 120 inward pointing main PMTs. The CCM120 experiment met expectations and performed a sensitive search for sub-GeV dark matter via coherent nuclear scattering with $1.79\times10^{21}$ Protons On Target (POT)~\cite{CCM:2021leg,CCM:2021yzc}. Due to the intense scintillation light production and short 14\,cm radiation length in LAr \cite{LArproperties}, the relatively large CCM detector has good response to electromagnetic signal events in the energy range from 100\,keV up to 100's of MeV.  

Another key feature of CCM is that it uses fast beam and detector timing to isolate prompt ultra-relativistic particles originating in the target. This can distinguish signal from the significantly slower neutron backgrounds that arrive approximately 225\,ns after the start of the beam pulse (relativistic particles traverse the 23\,m distance in 76.6\,ns) \cite{CCM:2021leg}. Furthermore, the Lujan beam low duty factor of $\sim 10^{-5}$ and extensive shielding are efficient at rejecting steady state backgrounds from cosmic rays, neutron activation, and internal radioactivity from PMTs and $^{39}$Ar.   

In order to determine the sensitivity reach of CCM's ongoing run, we use the  beam-on background distribution determined from the recent CCM120 run \cite{CCM:2021leg}, with a further expected factor of 100 reduction from extensive improvements in shielding, veto rejection, energy and spatial resolution, particle identification analysis, and reduced beam width.  Further details of the signal gamma-ray and electron event reconstruction and background rejection analysis is detailed in the recent CCM120 ALP search  \cite{CCM:2021jmk}, which share many similarities.

\begin{figure}[h]
    \centering
    \includegraphics[width=0.5\textwidth]{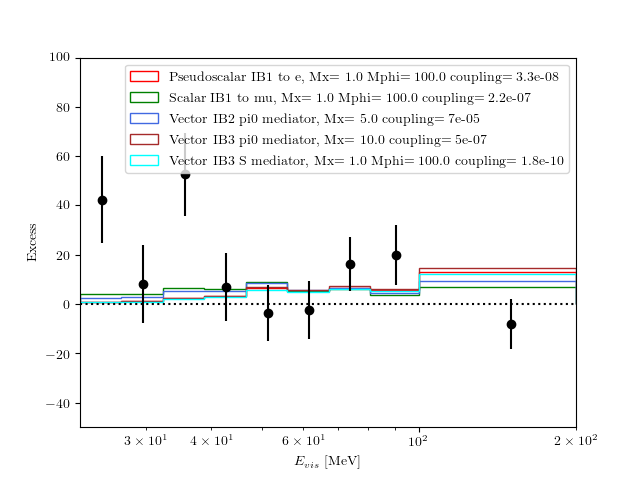}
    \caption{The CCM120 high energy subtraction spectrum used for the search search. Only events between 22.4 and 200 MeV reconstructed energy were included in this spectrum. Also shown are example spectra from the Dark Sector for the 5 models tested with CCM120, using points on the 68\% confidence level.}
    \label{fig:CCM120signal}
\end{figure}

Since our MiniBooNE excess explanation requires dominant contributions from the charged pion decay (otherwise the data in the beam-dump mode measurement would rule it out),  the constraint for this parameter space mostly emerges from the elastic and inelastic analyses. The visible energy distribution for the events at CCM120 is in the range 10-70 MeV (as shown in Fig.~\ref{fig:CCM120signal}) for various scenarios described in \S~\ref{sec:models}. In Fig.~\ref{fig:mb_chi2}, we show the allowed parameter space where the MiniBooNE excess can be explained after satisfying the LSND constraints, in addition to the comparison with projected sensitivities for CCM assuming the null hypothesis.

\section{Results}
\label{sec:results}

\begin{figure*}[ht]
    \centering
    \includegraphics[width=0.325\textwidth]{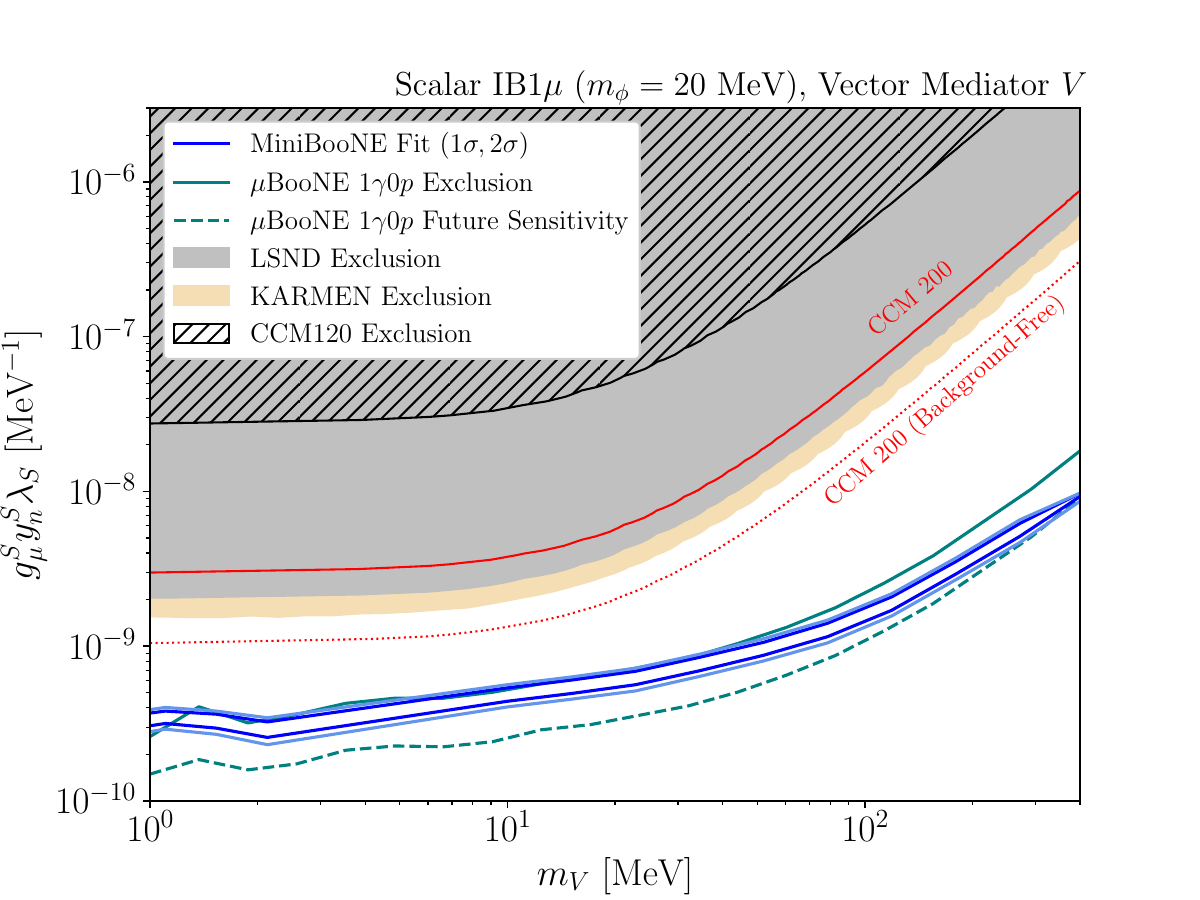}
     \includegraphics[width=0.325\textwidth]{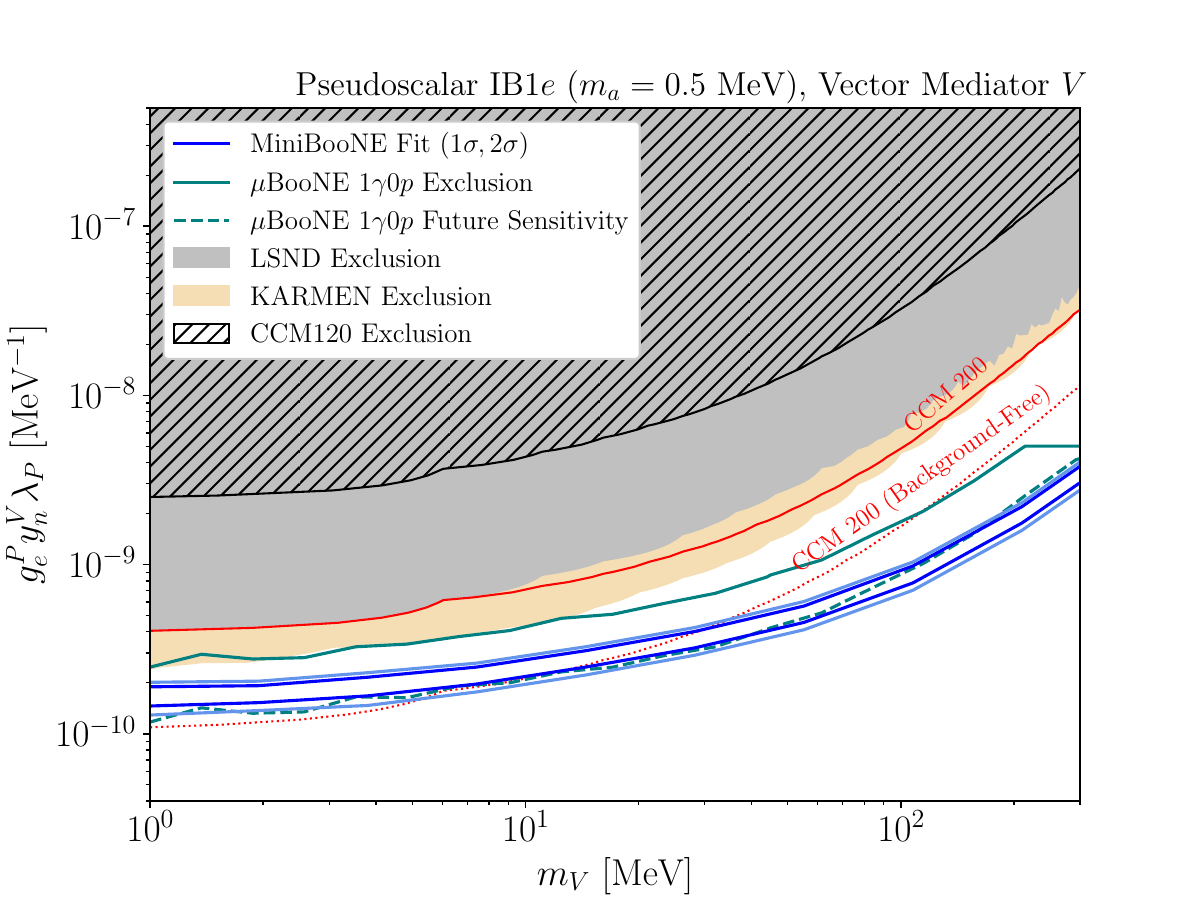}
      \includegraphics[width=0.325\textwidth]{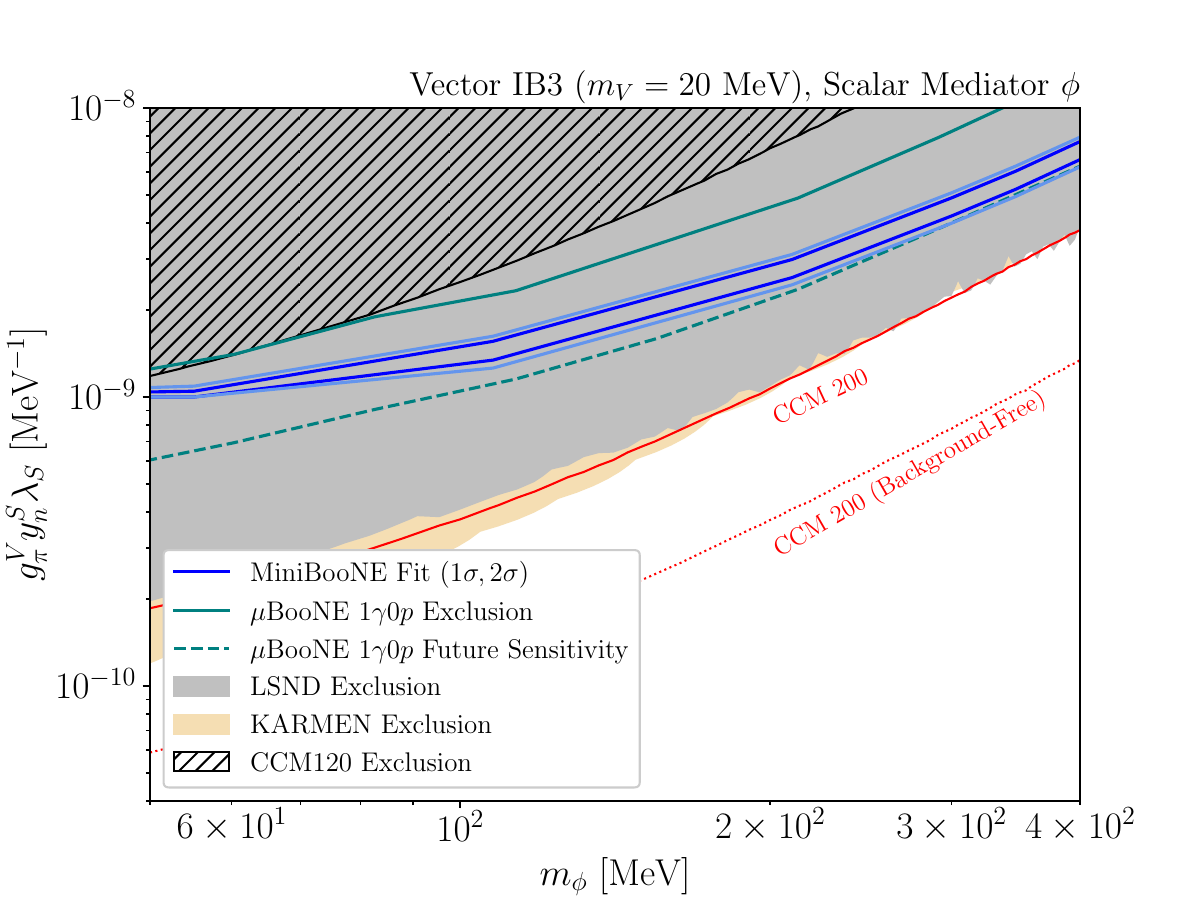}
    \caption{Parameter space for the two-mediator models, consisting of one long-lived boson produced through charged pion three-body decays and scattering via a secondary heavy mediator are shown. Left: a 20 MeV scalar $\phi$ coupling to muons and scattering via a vector mediator $V$ as in Eq.~\ref{eq:IB1muS}. Center: a 0.5 MeV pseudoscalar $a$ coupling to electrons and scattering via a vector mediator $V$ as in Eq.~\ref{eq:IB1eP}. Right: a 20 MeV vector $V$ coupling through the IB3 pion contact interaction and scattering via a massive scalar $\phi$ as in Eq.~\ref{eq:IB3VMed}.  The exclusions (CCM120, KARMEN, LSND, and MicroBooNE) and projections  (CCM200 and MicroBooNE) are shown at 95\% C.L., while the MiniBooNE fits are shown at 68\% and 95\% C.L. in dark and light blue, respectively.}
    \label{fig:mb_chi2}
\end{figure*}
In Fig.~\ref{fig:mb_chi2} we show the resulting constraints set by the CCM120 data, projections for CCM200 along with the preferred MiniBooNE regions (at 1$\sigma \simeq 68$\% and 2$\sigma \simeq 95$\% C.L.), and the constraints from the LSND DIF and DAR analyses (see Table~\ref{tab:lsnd}) for three possible decay models and scattering model scenarios. These models are named according to (i) the type of long-lived boson and decay mode (see Fig.~\ref{fig:3body}), e.g. Scalar IB1$\mu$ to indicate the IB1 decay channel through a coupling to the muon leg, and (ii) the type of mediator (scalar, pseudoscalar, or vector) used in the scattering via the interactions in Fig.~\ref{fig:dark_primakoff}. All curves correspond to 95\% C.L. Also shown are the limits extracted from the MicroBooNE 
$1{\gamma}0p$ data~\cite{MicroBooNE:2021zai}. In addition, we calculate a projected sensitivity for an exclusive $1\gamma 0 p$ search at MicroBooNE with greater signal efficiency (we assume a 3 times multiplied efficiency profile to an average of $15$\%~\cite{MicroBooNE:2022oim}) and more exposure ($1.11 \times 10^{21}$ POT~\cite{osti_2405957}) by scaling the background event distribution from ref.~\cite{MicroBooNE:2021zai} accordingly and testing a null-result $\Delta \chi^2$.

Beginning with Fig.~\ref{fig:mb_chi2} (left), we consider the parameter space for a long-lived scalar particle $\phi$ produced via the IB1 decay $\pi^\pm \to \mu \nu \phi$ through a muonic coupling, and scattering through $\phi N \to \gamma N$ photoconversion via a massive vector mediator $V$ (see also \S~\ref{sec:models} and \S~\ref{sec:pheno} for details). The decay and scattering are described by the phenomenological Lagrangian
\begin{equation}
\label{eq:IB1muS}
    \mathcal{L}_{int} \supset g_\mu^S \phi \bar{\mu}\mu + y_n^V V_\mu \bar{N} \gamma^\mu N - \frac{\lambda_S}{4} \phi F_{\mu\nu} H^{\mu\nu}\, + \textrm{h.c.}
\end{equation}
with $H_{\mu\nu}\equiv \partial_\mu V_\nu - \partial_\nu V_\mu$ and vector mass $m_V > m_\phi$. The event rate is proportional to the coupling product $g_\mu^S y_n^V \lambda_S$.  For this setup, we can fix the mass of the long-lived scalar to 20 MeV and vary $m_V$, for which we find that the fit to the MiniBooNE target and dump mode data (blue) lies around the scale  $g_\mu^S y_n^V \lambda_S \simeq 10^{-9}$ MeV$^{-1}$ at the 1$\sigma$ and 2$\sigma$ levels. The black hatched region is constrained by the CCM120 data, while constraints by LSND shown in olive are more stringent but do not rule out any of the preferred parameter space from the MiniBooNE fit. Conversely, MicroBooNE's $1\gamma 0p$ data (solid teal) excludes more parameter space up to about a factor of 2 larger in the coupling product across all mediator masses, and we see that with a dedicated single photon search (dashed teal) and more exposure, MicroBooNE will be able to test the preferred model parameter space to explain the MiniBooNE excess. We also expect that the other SBN experiments, SBND and ICARUS, should also be very sensitive to this parameter space beyond the MiniBooNE preferred region.

In Fig.~\ref{fig:mb_chi2} (center), the parameter space for a long-lived pseudoscalar coupling to electrons and produced through IB1 $\pi^\pm \to e \nu a$ decays is shown as a function of the coupling product and mass of a vector mediator taking place in the $a N \to \gamma N$ scattering via similar interactions,
\begin{equation}
\label{eq:IB1eP}
    \mathcal{L}_{int} \supset -ig_e^P a \bar{\mu}\gamma^5 \mu + y_n^V V_\mu \bar{N} \gamma^\mu N - \frac{\lambda_P}{4} a F_{\mu\nu} \Tilde{H}^{\mu\nu}\, + \textrm{h.c.}
\end{equation}
We fix the pseudoscalar mass $m_a < 2 m_e \simeq 1$ MeV, otherwise $a \to e^+ e^-$ decays would be kinematically allowed and may be incompatible with the excess signal, and again take $m_V > m_a$. In this scenario, the event rate is proportional to the coupling product $g_e^P y_n^V \lambda_P$, and a result similar to Fig.~\ref{fig:mb_chi2}, left is found where the entirety of the MiniBooNE preferred region is allowed by the existing constraints and in reach of the future MicroBooNE $1\gamma 0 p$ search.

In Fig.~\ref{fig:mb_chi2} (right) we consider a third scenario in which a massive vector mediator is long-lived and couples to the charged pion through a contact interactions, and subsequently scatters via a massive scalar mediator with mass $m_\phi > m_V$. We take the effective interaction Lagrangian
\begin{align}
\label{eq:IB3VMed}
    \mathcal{L}_{int} \supset& \, y_n^S \phi \overline{N} N - \frac{\lambda_S}{4} a F_{\mu\nu} \Tilde{H}^{\mu\nu}  \nonumber \\
    &  -ig_\pi^V \pi^+ \bar{\mu}\gamma^\rho (1-\gamma^5) \nu V_\rho \,+\, \textrm{h.c.}
\end{align}
However, in this case we find that the favored parameter space to explain the MiniBooNE excess is largely ruled out at 95\% C.L. by LSND and KARMEN, in part due to the characteristically smaller cross section at higher energies for coherent $V N \to \gamma N$ scattering via the heavy scalar, giving the experiments with lower energy beams an enhanced detection sensitivity.

\begin{figure*}[ht]
    \centering
       \includegraphics[width=0.325\textwidth]{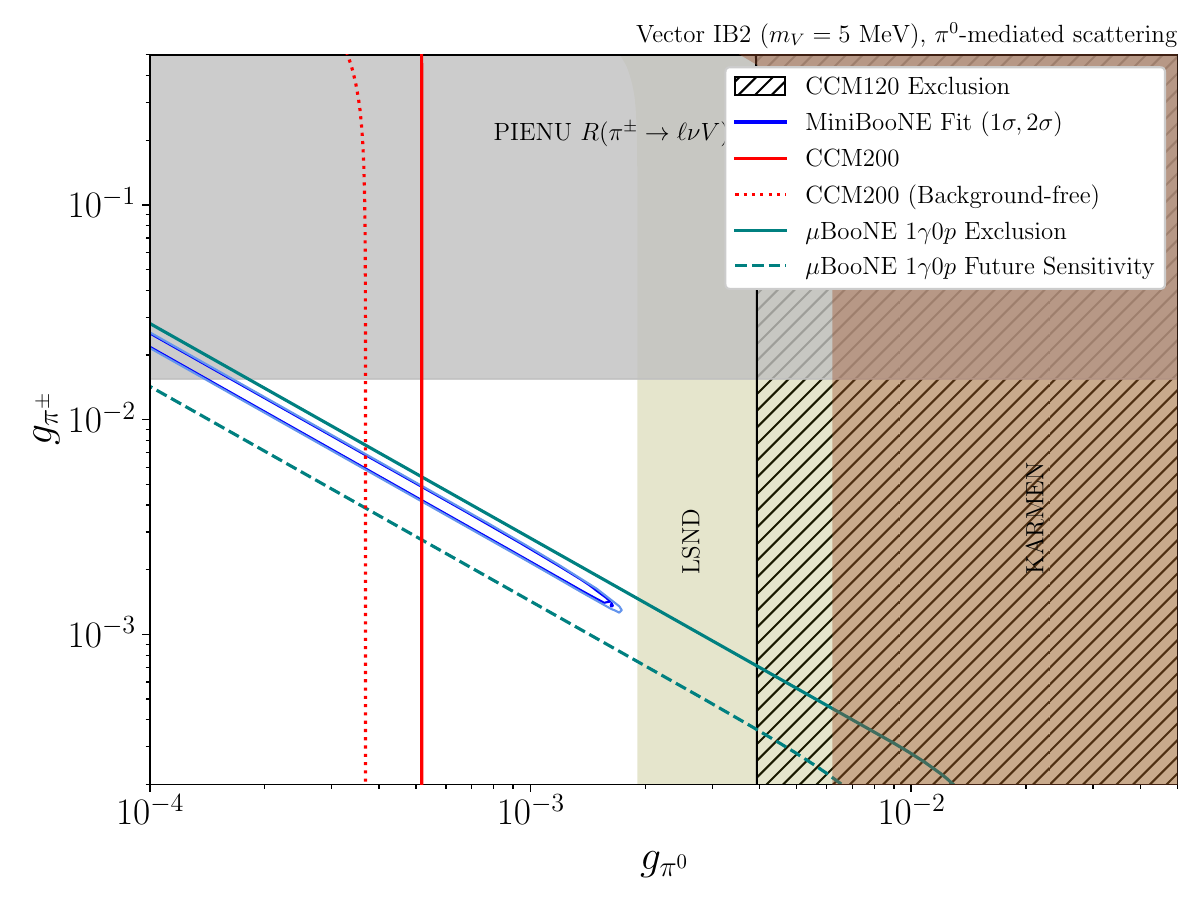}
       \includegraphics[width=0.325\textwidth]{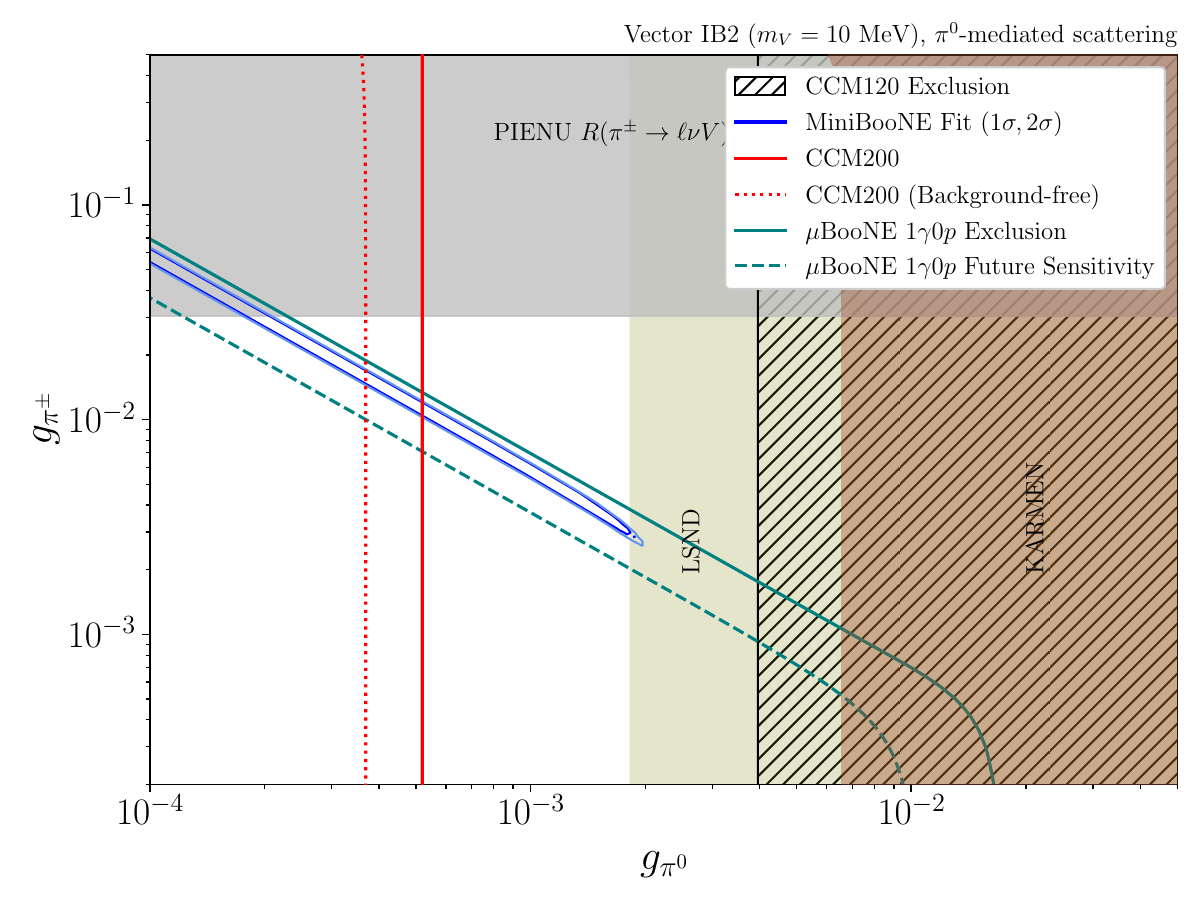}
       \includegraphics[width=0.325\textwidth]{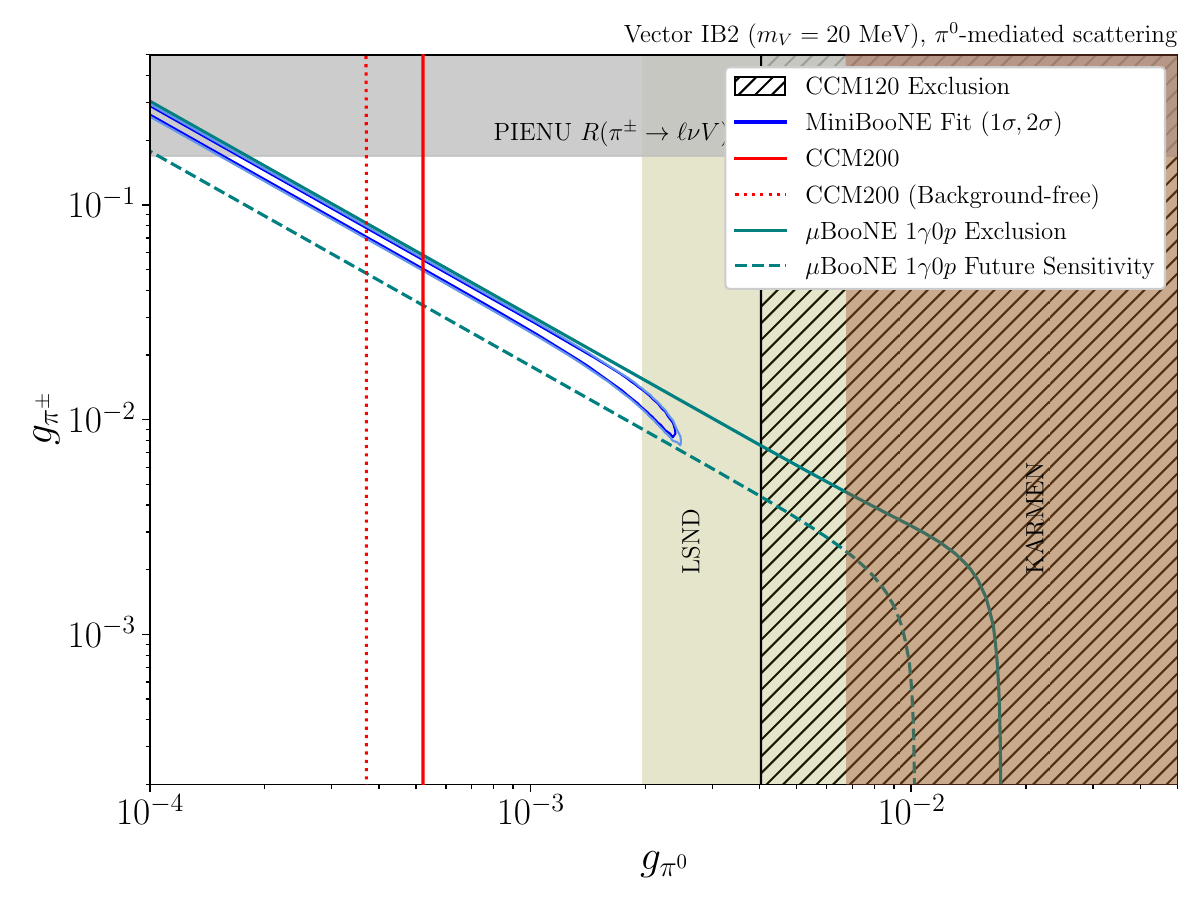}
    \caption{Parameter space for the single mediator scenario where a massive vector $V$ couples to the pion doublet via charged pion coupling $g_{\pi^\pm}$ and neutral pion coupling $g_{\pi^0}$ as in Eq.~\ref{eq:singlemedLag}. The production channels via these couplings are therefore neutral pion decay $\pi^0 \to \gamma V$ and IB2 decay $\pi^\pm \to \ell \nu V$, while the detection takes place via $\pi^0$-mediated $V N \to \gamma N$ scattering. We vary the vector mass $m_V$ from left to right as $5$ MeV (left), $10$ MeV (middle), and $20$ MeV (right).The exclusions (CCM120, KARMEN, LSND, and MicroBooNE) and projections (CCM200 and MicroBooNE) are shown at 95\% C.L., while the MiniBooNE fits are shown at 68\% and 95\% C.L. in dark and light blue, respectively.}
    \label{fig:single_med_limits}
\end{figure*}

In the second class of phenomenological model, we consider a single long-lived vector mediator that couples to quarks and enters the pion sector via the $\chi$-PT Lagrangian in Eq.~\ref{eq:singlemedLag};
\begin{align}
\label{eq:singlemedLag}
   \mathcal{L}_{int} \supset & \, ig_{\pi^\pm} V_\mu \pi^+ (\partial^\mu \pi^-) + g_{\pi^0} \frac{e}{16\pi f_\pi}\pi^0 F_{\mu\nu}\Tilde{H}^{\mu\nu} \nonumber \\
   &- ig_{\pi NN} \pi^0 \overline{N} \gamma^5 \tau_3 N
\end{align}
See again \S~\ref{sec:models} and \S~\ref{sec:pheno} for more details. In Fig.~\ref{fig:single_med_limits} we show the parameter space sensitivities and constraints for the IB2 decay model for $m_V = 5, 10, 20$ MeV. The CLs for the MiniBooNE fit are shown (blue) for the combination of $\nu$, $\bar{\nu}$, and beam-dump-mode runs, exclusions set by the CCM120 engineering run are shown by the black hatched region, and future sensitivity expected in CCM200 with upgrades (red). Also shown are constraints from LSND (light yellow), KARMEN (brown), and rare charged pion decay searches from PIENU (gray). In this case we have production channels from both charged pion decays ($\pi^\pm \to \ell \nu V$), for which we take the IB2 decay mode as a benchmark, and neutral pion decays ($\pi^0 \to V \gamma$). Constraints from $\pi^\pm$ decay width measurements can be directly applied to this parameter space in terms of $g_{\pi^\pm}$, and for these we take the bounds from PIENU~\cite{PIENU:2021clt} which have set constraints on invisible radiative decays $\pi^\pm \to e^\pm \nu X$ and $\pi^\pm \to \mu \nu^\pm X$ dependent on the $X$ mass.

Since the neutral pion decay channel is active in this scenario, it can dominate as a production channel to produce the LLP at the lower-energy beam target experiments (CCM, LSND, and KARMEN) where both the charged and neutral pions are isotropic. This contrasts with the case of MiniBooNE and MicroBooNE that take advantage of the magnetic focusing horns to enhance the fluxes driven from the charged pion decays. This results in a few qualitative differences, namely that at the lower-energy beam targets, the neutral pion decays will produce relatively more energetic LLPs than the charged pions (contrast this with the case in Fig.~\ref{fig:microboone_events}, for example). These neutral pion decay energies can be up to $\mathcal{O}(100)$ MeV, where the cross section of scattering is also larger (see Fig.~\ref{fig:xs}). These factors suppress the event rates at KARMEN, whose energy region of interest is limited below 40 MeV; hence, we see a more stringent exclusion from CCM120 relative to KARMEN in Fig.~\ref{fig:single_med_limits} in spite of KARMEN having more exposure and a larger detector size. On the other hand, LSND does set a more stringent exclusion, but it is also limited in reach due to the detector being only 12 degrees off-axis, capturing only the tail of the neutral pion momentum distribution which peaks instead at 90 degrees off-axis.

This range of parameter space preferred by the MiniBooNE fit in Fig.~\ref{fig:single_med_limits} is not yet excluded by LSND at 95\% C.L., and within reach of a future $1\gamma 0 p$ search at MicroBooNE. In addition, constraints from $\pi^0$ decay width measurements apply to the $g_{\pi^0}$ coupling in this parameter space, but they apply only to larger values of the coupling that are not shown in the plot. The constraints from PIENU also begin to relax while moving to larger masses, due to the weaker branching ratio of $\pi^+\pm \to \ell \nu V$ with higher mass (see e.g. Fig.~\ref{fig:brs_vector}). In each case, we again find that a $1\gamma 0 p$ search at MicroBooNE with higher exposure should be sensitive to this explanation of the MiniBooNE excess at 95\% C.L.

While we have only shown three mass points here, these trends are expected to hold through to the kinematic upper limit of the kinematically allowed vector boson mass in the 2-body and 3-body decays, $m_V \lesssim m_\pi$. For lower masses, the landscape of constraints relative to the MiniBooNE fit will be very similar to the case in Fig.~\ref{fig:single_med_limits}, left, except globally shifted to lower $g_{\pi^\pm}$ couplings to compensate for the larger branching ratios at lower masses. The relevant constraint that should apply at lower masses would be from cosmological considerations; we generically expect a limit on the mass $m_V \lesssim 10$ MeV due to its impact on $\Delta N_{eff}$~\cite{Escudero:2018mvt}, which is, however, model-dependent. 

\section{Conclusions}
\label{sec:conclude}
The rare three-body decays of charged pions and kaons to new states in the MeV mass scale as a dark-sector explanation for the MiniBooNE excess has shed light on a potential correlation between the mesonic sector and new physics solutions. By investigating an extended range of phenomenological models, we have demonstrated that these scenarios, incorporating long-lived particles generated in the three-body decays of charged mesons and two-body anomalous decays of neutral mesons, can be consistent with constraints from LSND, KARMEN, and MicroBooNE experiments. In particular, we found that in the context of these models with a long-lived particle and a heavy mediator to facilitate photoconversion scattering, the MiniBooNE excess data preferred a mediator in the mass range $m_Y \gtrsim 10-100$ MeV. In all cases, scattering with detector atomic nuclei was considered, so it may be interesting to probe new mediators in this mass range with generalized hadronic couplings in separate experiments. Secondly, the inelastic nuclear responses to the mediators we have considered is an interesting possibility to study, namely in the context of the LSND excess which we have set aside for the time being. One could also examine the same inelastic channels as they contribute to the event spectra at KARMEN, MiniBooNE, MicroBooNE, and CCM, although this requires a detailed shell-model description of the nucleus coupled to the new mediators we have used.

The forthcoming analysis of the current CCM200 data taking campaign will have the ability to test dark-sector explanations to the MiniBooNE excess, especially for new long-lived particles coupled to the pion doublet; as a stopped-pion experiment, it can leverage the neutral pion production and its close proximity to the proton beam target. In this way, stopped-pion experiments have more sensitivity via the neutral pion channels to probe this set of solutions in a complementary way to short baseline experiments, whose magnetic horns produce instead a focused flux of charged mesons. Long-lived vector mediators that couple to the pion doublet around 5 MeV in mass, as preferred by the fits to MiniBooNE data, are now susceptible to searches through both stopped-pion experiments as well as rare meson decay searches. Though not within the scope of this work, there is no reason to not expand the dark sector couplings to the meson octet which would include kaons, or to the broader hadronic spectrum of baryons and vector mesons. This analysis motivates such cases through the advantage of correlated couplings which open up multiple production and detection channels to constrain, and hopefully discover, solutions to anomalies in this fashion.\\

\section*{Acknowledgement}
We thank Michael Shaevitz and William Seligman for the dedicated programming work and feedback on the \texttt{modelB} routines for particle transport in the BNB horn system. We also thank Wooyoung Jang for the simulation work on the MiniBooNE pion fluxes and Noemi Rocco for the useful discussions. AAA-A, JCD and MC-E acknowledge support from DGAPA-UNAM grant No. PAPIIT-IN104723. We acknowledge the support of the Department of Energy Office of Science, Los Alamos National Laboratory LDRD funding, and funding from the National Laboratories Office at Texas A\&M. We acknowledge that portions of this research were conducted with the advanced computing resources provided by Texas A\&M High Performance Research Computing. We also wish to acknowledge support from the LANSCE Lujan Center and LANL's Accelerator Operations and Technology (AOT) division.  This research used resources provided by the Los Alamos National Laboratory Institutional Computing Program, which is supported by the U.S. Department of Energy National Nuclear Security Administration under Contract No.\,89233218CNA000001.

\bibliography{main}

%%% APPENDIX %%%

{\onecolumngrid
\appendix

\section{Meson Flux Simulations at the BNB}
\label{app:horn}
To simulate the focused charged meson decays that take place in the BNB horn system, we begin by simulating the proton beam spot that sources the charged mesons, shown in Fig.~\ref{fig:beamspot}, based on the normal distribution of protons given in~\cite{Schmitz:2008zz} which source a $\pi^\pm$ at beam spot position $(x,y)$ and depth $z$ into the target given from the interaction probability $1 - e^{-\sigma(p) n z}$ based on the pion production cross section in Eq.~\ref{eq:pion_xs} and Be density $n = 1.85$ g/cm$^3$. The proton momenta are also generated by a parameterization.
\begin{figure}[ht]
    \centering
    \includegraphics[width=0.41\textwidth]{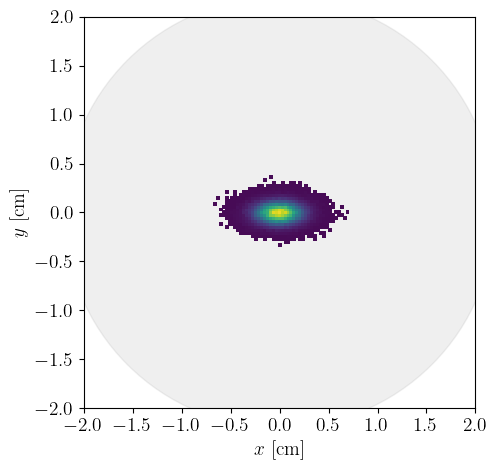}
    \caption{Simulation of the BNB beam spot distribution of protons on target as a function of $(x,y)$ over the face of the Be target, shown in gray.}
    \label{fig:beamspot}
\end{figure}
We use this beam spot to generate a monte carlo sample of pion production vertices. Their momenta and production angles with respect to the progenitor proton direction can be expressed using the Sanford-Wang parameterization given in ref.~\cite{MiniBooNE:2008hfu}, shown in Fig.~\ref{fig:mb_pion_fluxes}. This scheme parameterizes the total pion production cross section as follows;
\begin{equation}
    \dfrac{\partial^2\sigma(p + \textrm{Be} \to \pi^\pm + X)}{\partial p \partial\Omega} = c_1 p^{c_2} \bigg(1 - \frac{p}{p_B - c_9}\bigg) \exp\bigg(-c_3 \frac{p^{c_4}}{p_B^{c_5}} - c_6 \theta (p - c_7 p_B (\cos\theta)^{c_8} \bigg)
\end{equation}
where the constants associated with $\pi^+$ ($\pi^-$) production are repeated in Table~\ref{tab:sw} for convenience.
\begin{table}[h]
    \centering
    \begin{tabular}{|c|c|c|c|c|c|c|c|c|c|}
    \hline
         Type & $c_1$ & $c_2$ & $c_3$ & $c_4$ & $c_5$ & $c_6$ & $c_7$ & $c_8$ & $c_9$  \\
         \hline
         $\pi^+$ & 220.7 & 1.080 & 1.0 & 1.978 & 1.32 & 5.572 & 0.0868 & 9.686 & 1.0  \\
         \hline
         $\pi^-$ & 213.7 & 0.9379 & 5.454 & 1.210 & 1.284 & 4.781 & 0.07338 & 8.329 & 1.0 \\
         \hline
    \end{tabular}
    \caption{Sanford-Wang cross section parameters at the BNB target.}
    \label{tab:sw}
\end{table}
while the total cross section is parameterized as
\begin{equation}
\label{eq:pion_xs}
    \sigma(p) = a + b p^n + c(\ln(p))^2 + d \ln(p)
\end{equation}
with
\begin{align}
    a &= 307.8 \nonumber \\
b &= 0.897 \nonumber \\
c &= -2.598 \nonumber \\
d &= -4.973 \nonumber \\
n &= 0.003
\end{align}

Taking the position $(x, y, z)$ from the beam spot simulation and the momentum, polar angle, and azimuthal angle $(p, \cos\theta, \phi)$ from a weighted MC simulation of the Sanford-Wang cross section, the pion flux is prepared for simulated transport through the remainder of the beam target and horn system. For this we use a simple geometric model of the BNB horn shape and magnetic field profile as inputs to a Runge-Kutta charged particle transport routine~\cite{RKHorn2024}. Some sample trajectories are shown in Fig.~\ref{fig:horn_sim}.

\begin{figure}
    \centering
    \includegraphics[width=0.45\textwidth]{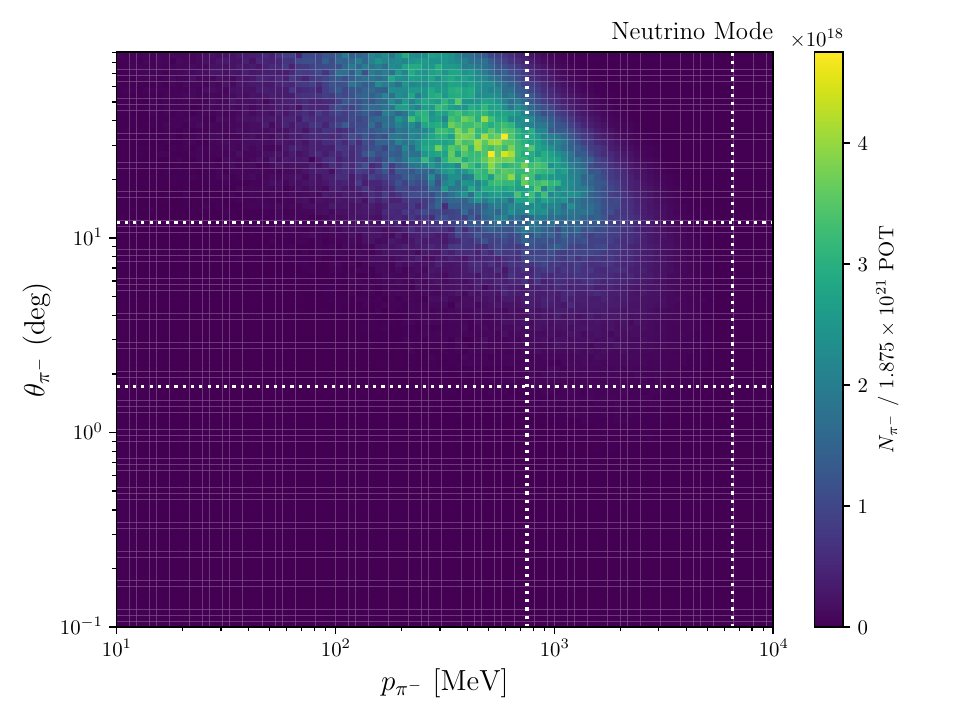}
    \includegraphics[width=0.45\textwidth]{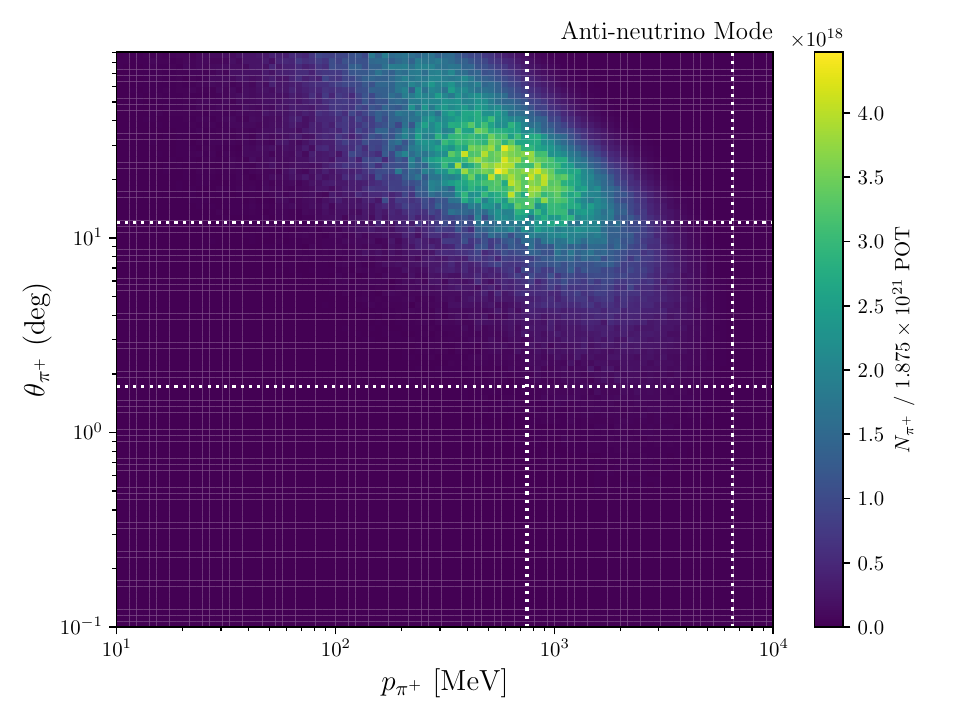}
    \caption{Charged pion fluxes produced at the BNB via a monte carlo treatment of the Sanford-Wang parameterization of the charged pion production cross section.}
    \label{fig:mb_pion_fluxes}
\end{figure}

\begin{figure}
    \centering
    \includegraphics[width=0.6\textwidth]{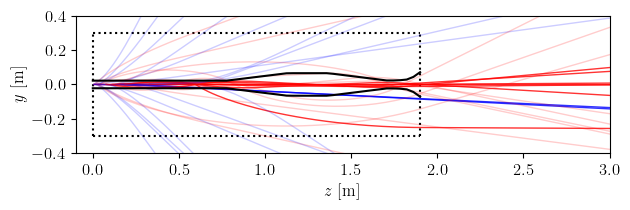}
    \caption{Simulation of the BNB horn with example trajectories showing the $\pi^+$ (red) and $\pi^-$ (blue) transport in the horn system for FHC polarization.}
    \label{fig:horn_sim}
\end{figure}

The post-horn flux distribution using $5\times 10^5$ simulated POT is shown in Fig.~\ref{fig:pre_post_horn_flux} as a function of pion angle with respect to the beam axis. For comparison, the equivalent detector solid angle coverages of MicroBooNE and MiniBooNE are 1.2 mrad and 3 mrad, respectively. We validate this distribution in a pragmatic way by checking that it predicts a neutrino spectrum at the MiniBooNE detector that is consistent with what is reported by the collaboration. To do this, we take the focused $\pi^\pm$ fluxes predicted by Sanford-Wang and perform a 2-body decay monte carlo algorithm on the charged pions, allowing them to decay to $\nu_\mu$, $\mu$ at some distance $x$ away from the production site in the target, where $x$ itself is drawn from a distribution like $\exp(-x/v\tau_{\pi^\pm})$ using the pion lifetimes.

\begin{figure}
    \centering
    \includegraphics[width=0.45\textwidth]{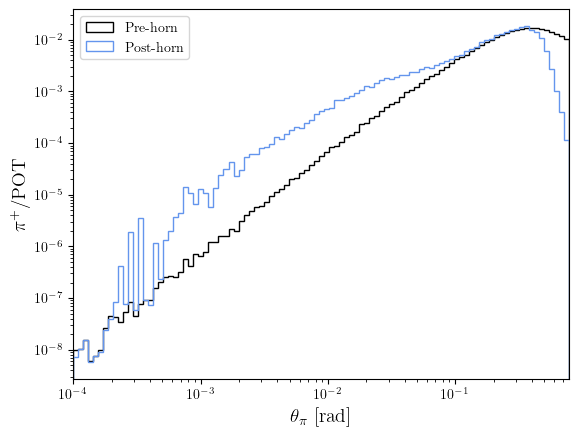}
    \caption{The pre-horn $\pi^+$ flux (black) using the Sanford-Wang parameterized cross section convolved with the BNB proton beamspot, and the post-horn flux (blue) after modeling the transport through the magnetic horn system are both shown as a function of the pion angle with respect to the beam axis. In both cases $5\times 10^5$ simulated POT were used.}
    \label{fig:pre_post_horn_flux}
\end{figure}

Additionally, the distributions for neutral $\pi^0$ that are produced in the BNB target and the dump are shown in Fig.~\ref{fig:dump_fluxes}. The differences in the $\pi^0$ energy and angle distributions at the target versus the dump can be attributed to the larger size of the dump and the differences in material on which the beam impinges. To simulate events from the 2-body decays of $\pi^0 \to \gamma X$, we perform the decay simulation in the pion rest frame and boost $X$ into the lab frame. Since $X$ is long-lived and weakly coupled, it can be invisibly transported to the detector. Simultaneously we again check that the $X$ production angle with respect to the beam line is within the detector solid angle. 

The $\pi^0$ kinematic distributions at the MiniBooNE dump and target are shown in Fig.~\ref{fig:dump_fluxes}, generated from GEANT4 simulation. In each case the rates on the color bar are normalized to $10^4$ simulated POT. The discrepancy between the two fluxes, which are $\sim \mathcal{O}(10)$ larger in rate and more energetic in the target mode, can be explained by the choice of material; the target material (Be) has a much lighter nucleus than the dump (steel). Neutral pions produced from protons impinging on the nuclei are long-lived on nuclear scales -- $\tau c \simeq 25$ nm for boosted pions -- and will undergo multiple scattering and absorption much more often in a heavy nucleus than a light one.

\begin{figure}
    \centering
    \includegraphics[width=0.4\textwidth]{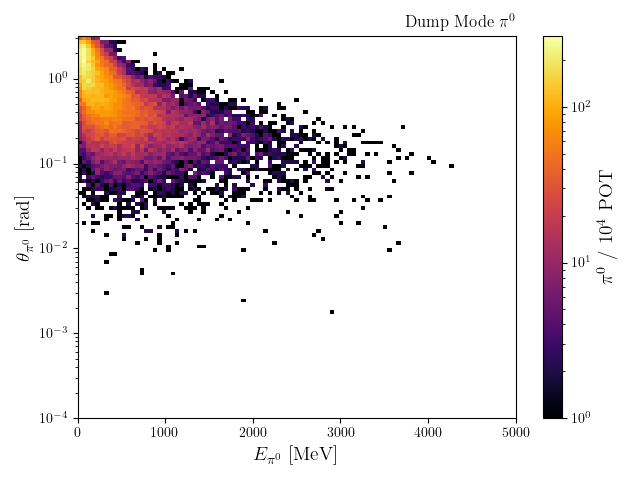}
    \includegraphics[width=0.4\textwidth]{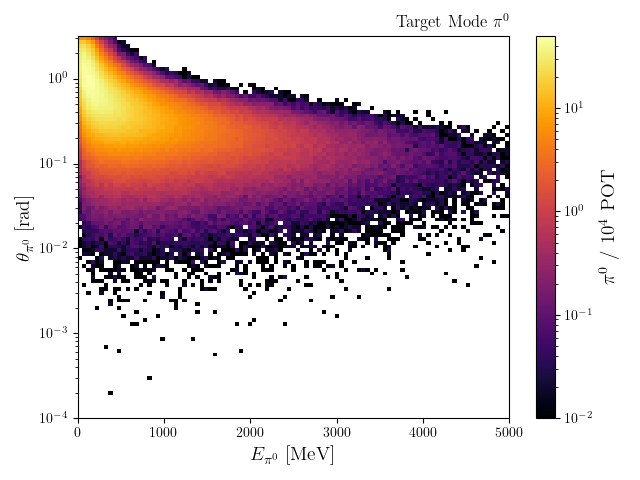}
    \caption{Beam-dump-mode and target-mode $\pi^0$ fluxes, generated from GEANT4 simulation, at the MiniBooNE dump (left) and BNB target (right) as a function of their kinetic energy and angle of travel with respect to the beam axis.}
    \label{fig:dump_fluxes}
\end{figure}

\section{Treatment of 3-body Decay Kinematics}

For the charged meson three-body decay $M(P) \to \ell(p_1) \nu(p_2) a(p_3)$, we make use of the Dalitz variables $m_{ij}^2 = (p_i + p_j)^2$. In the lab frame, we have
\begin{align}
    m_{12}^2 &= (p_1 + p_2)^2 = (P - p_3)^2 = M^2 - 2ME_a + m_a^2 \\
    m_{23}^2 &= (p_2 + p_3)^2 = (P - p_1)^2 = M^2 - 2ME_\ell + m_\ell^2 \\
    m_{13}^2 &= (p_1 + p_3)^2 = (P - p_2)^2 = M^2 - 2ME_\nu \\
    m_{13}^2 &= M^2 + m_\ell^2 + m_a^2 - m_{12}^2 - m_{23}^2 .
\end{align}
This set of variables allows us to write
\begin{equation}
    d\Gamma = \dfrac{1}{(2\pi)^3 32 M^3} \braket{|M|^2} dm_{23}^2 dm_{12}^2
\end{equation}
and re-express $m_{12}^2$ in terms of $E_a$, since $|dm_{12}^2| = 2MdE_a$, allowing us to integrate over $m_{23}^2$;
\begin{equation}
    \dfrac{d\Gamma}{dE_a} =  \int^{(m_{23}^2)_{max}}_{(m_{23}^2)_{min}} \dfrac{1}{(2\pi)^3 16 M^2} \braket{|M|^2} dm_{23}^2 .
\end{equation}
This has bounds
\begin{equation}
    (m_{23}^2)^{max}_{min} = (E_2^* + E_3^*)^2 - \bigg(E_2^* \mp \sqrt{{E_3^*}^2 - m_a^2}\, \bigg)
\end{equation}
with the starred energies defined as
\begin{align}
    E_2^* &= \dfrac{m_{12}^2 - m_\ell^2}{2 m_{12}}, \, \, \,  E_3^* = \dfrac{M^2 - m_{12}^2 - m_a^2}{2 m_{12}}
\end{align}
Finally, we can integrate over $E_a$ making use of the fact that $m_\ell^2 < m_{12}^2 < M^2 + m_a^2 - 2Mm_a$ to get the $E_a$ limits;
\begin{equation}
    m_a < E_a < \dfrac{M^2 + m_a^2 - m_\ell^2}{2 M}.
\end{equation}
Using this integration scheme, we show the total branching ratios for IB1, IB2, and IB3/contact decays, broken down by decay channel ($\ell = e$ or $\ell = \mu$), in Fig.~\ref{fig:brs_vector}.
\begin{figure}[h!]
    \centering
    \includegraphics[width=0.31\textwidth]{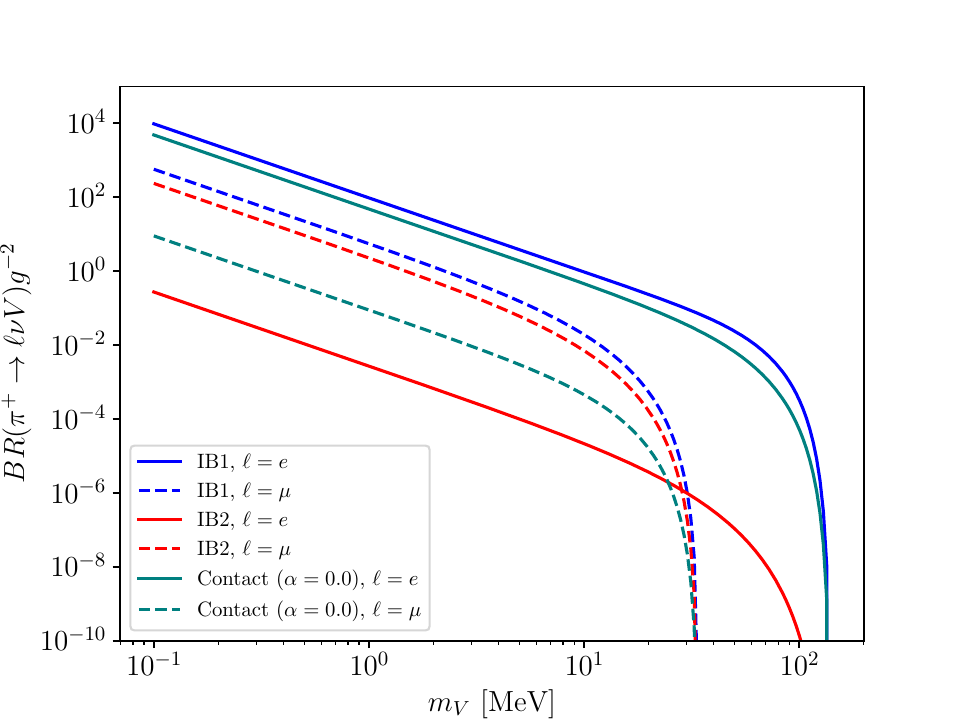}
    \includegraphics[width=0.31\textwidth]{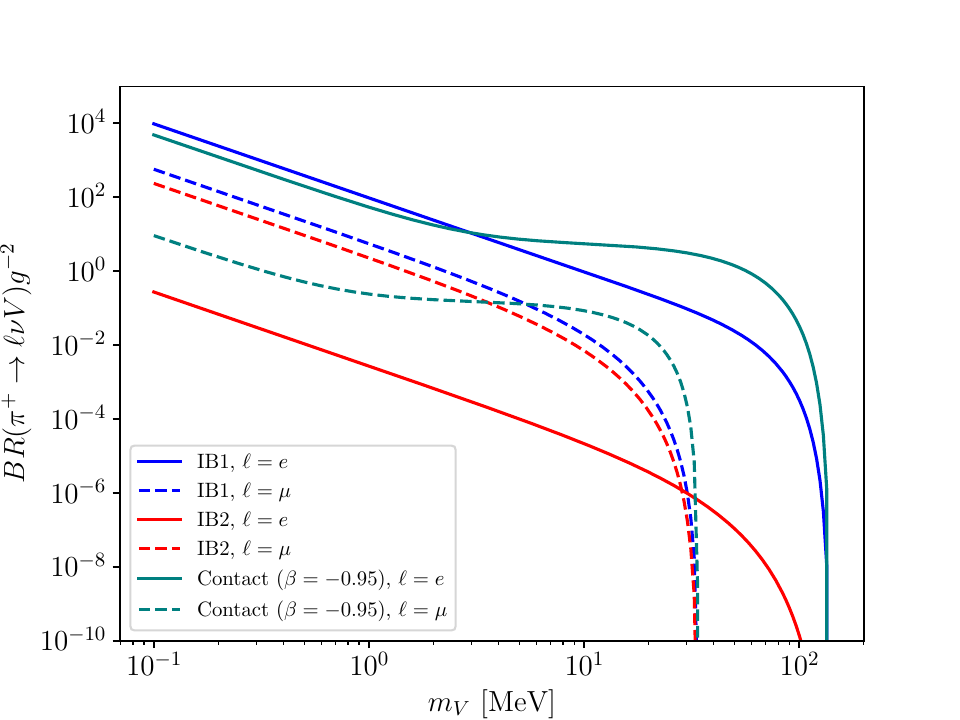}
    \includegraphics[width=0.31\textwidth]{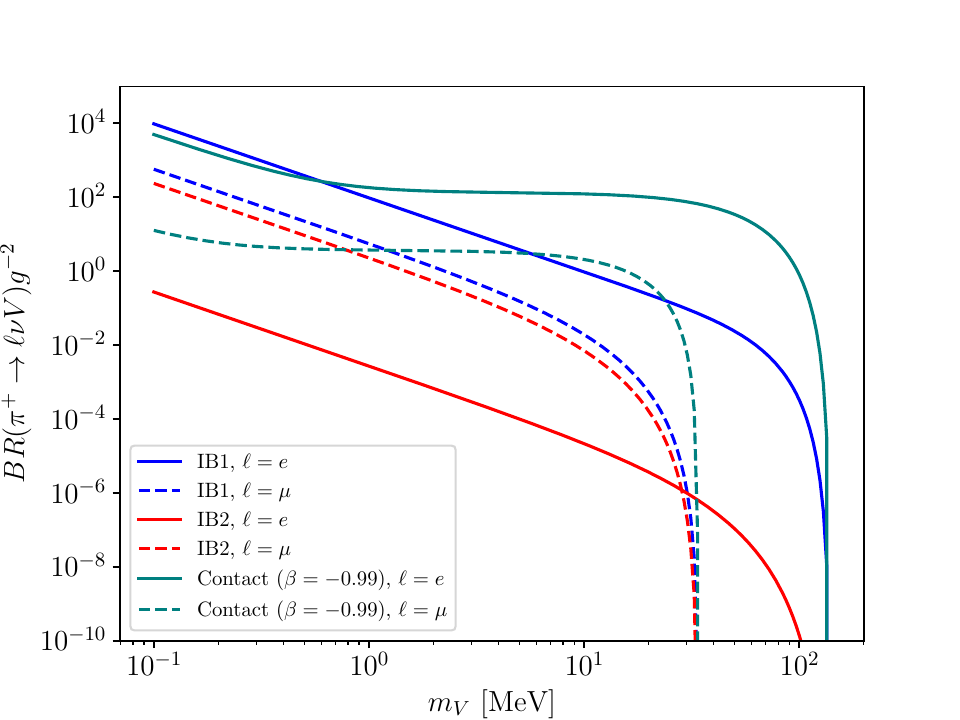}
    \caption{Branching Ratios for Vector IB2 and IB3 interactions with various choices of the coefficients of $\beta$ in Eq.~\ref{eq:IB3} with $\alpha=1.0$.}
    \label{fig:brs_vector}
\end{figure}
}

% ADDENDUM //////////////////////////////////////
% ADDENDUM //////////////////////////////////////
% ADDENDUM //////////////////////////////////////

\clearpage

\begin{center}
\large \textbf{Addendum to ``Testing Meson Portal Dark Sector Solutions to the MiniBooNE Anomaly at CCM''}\\
\end{center}
In this addendum, we extend the analysis to project sensitivity of upcoming MicroBooNE data to the long-lived particle models considered in the previous work. We find that a dedicated MicroBooNE analysis of the single photon final state with longer exposure and improved signal efficiency will be sensitive to these new physics explanations of the MiniBooNE excess, and could rule them out with a null observation at the 95\% confidence level.

\setcounter{section}{0} % Reset section counter

\section{MicroBooNE}
\label{sec:mub}

%\begin{figure}[ht!]
%    \centering
    %\includegraphics[width=0.5\textwidth]{plots/spectra/microboone_event_spectra.pdf}
    %\caption{Example event spectra prediction from 2-body decays of $\pi^0 \to \gamma V$ (blue) and 3-body decays of $\pi^+ \to \ell \nu V$ (red) at MicroBooNE in the 1$\gamma$0p topology. Here the LLP mass $m_V$ is fixed to 5 MeV. The MicroBooNE $1{\gamma}0p$ data and backgrounds are from~\cite{MicroBooNE:2021zai}.}
   % \label{fig:microboone_events}
%\end{figure}

The MicroBooNE collaboration performed an analysis of $\Delta \to N \gamma$ resonant production from a $\braket{0.8}$ GeV neutrino beam in several final state topologies, namely those with one photon and no hadronic activity ($1\gamma 0 p$) and one photon and one proton ($1\gamma 1 p$)~\cite{MicroBooNE:2021zai}, using data collected from $6.80 \times 10^{20}$ protons-on-target (POT) from the booster neutrino beam (BNB) source. The anomalous observation of excess electron-like events by MiniBooNE~\cite{MiniBooNE:2008yuf,MiniBooNE:2018esg,MiniBooNE:2020pnu} is a focus of existing and ongoing analyses of MicroBooNE data. In ref.~\cite{Aguilar-Arevalo:2023kvr}, building off of ref.~\cite{Dutta:2021cip}, a set of phenomenological models that could explain the MiniBooNE anomaly were studied, as well as the future sensitivity of the Coherent CAPTAIN Mills (CCM) experiment to these models was presented. These phenomenological models made use of production and detection mechanisms, schematically,
\begin{equation}
\underbrace{\begin{tikzpicture}
        \begin{feynman}
            \vertex (a) {\(\pi^\pm\)};
            \vertex [right=1.2cm of a, blob] (b) {\(\, \, \, \, \, \)};
            \vertex [right=1.2cm of b] (f2) {\(X\)};
            %\vertex [right=1.0cm of f2] (p) {\(+\)};
            \vertex [below left=0.3cm of b] (xl);
            \vertex [above right=1.2cm of b] (f1) {\(\nu\)};
            \vertex [below right=1.2cm of b] (f3) {\(\ell\)};
            
            \diagram* {
                (a) -- [scalar] (b) -- [fermion] (f1),
                (b) -- [anti fermion] (f3),
                (b) -- (f2)
            };
        \end{feynman}
    \end{tikzpicture}}_\textrm{Beam Target Production}
    \begin{tikzpicture}
           \begin{feynman}
                    \vertex (o3) {\(\)};
                    \vertex [above=1.0cm of o3] (p) {\(+\)};
           \end{feynman}
    \end{tikzpicture}
    \underbrace{\begin{tikzpicture}
    \begin{feynman}
         \vertex (o1);
         \vertex [above left=0.7cm of o1] (f1) {\(X\)};
         \vertex [above right=0.7cm of o1] (i1){\(\gamma\)};
         \vertex [below=0.7cm of o1] (o2);
         \vertex [right=0.7cm of o2] (f2) {\(N\)};
         \vertex [left=0.7cm of o2] (i2) {\(N\)};

         \diagram* {
           (i1) -- [boson] (o1) -- (f1),
           (o1) -- [edge label={\(Y\)}] (o2),
           (i2) -- [fermion] (o2),
           (o2) -- [ fermion] (f2),
         };
        \end{feynman}
       \end{tikzpicture}}_\textrm{Detection}
      \nonumber
\end{equation}
where a long-lived particle $X$ can be produced via 3-body decays of the charged mesons that are produced by the BNB target, propagating to the detector with negligible attenuation in flux, and scatter by exchanging a mediator $Y$ (arising from a dimension-5 operator) with nucleons in detector material. A secondary $X$ production channel from 2-body $\pi^0$ decays was also considered in some variations of the phenomenological models. In this addendum, we extend this analysis to forecast an expected event rate in future MicroBooNE data based off the preferred parameter space to explain the MiniBooNE anomaly with the various phenomenological models investigated in ref.~\cite{Aguilar-Arevalo:2023kvr}.

We calculate the expected event rate at MicroBooNE, again employing the simulation procedure for the charged and neutral pions produced in the BNB target and focused through the horn system working in neutrino-mode polarization used in ref.~\cite{Aguilar-Arevalo:2023kvr}. This procedure is exactly the same as before, integrating the new physics differential decay width for $\pi^+ \to X \ell \nu$, convolved with the simulated pion flux, over the solid angle spanned by the MicroBooNE detector's geometric cross section. To account for future improvements in the signal efficiency for an exclusive single-photon search, we assume an improvement by an \textit{ad hoc} factor of three multiplied by the $\sim5.29$\% signal efficiency from the first $1\gamma 0$p analysis, and a greater exposure of $1.11 \times 10^{21}$ POT as a benchmark future dataset; these assumptions are motivated in part by ref.~\cite{osti_2405957}. We then scale the background event distribution from ref.~\cite{MicroBooNE:2021zai} accordingly, but keep the same systematic uncertainties.

\section{Model Parameter Space}

\begin{figure}[h!]
    \centering
    \includegraphics[width=0.5\textwidth]{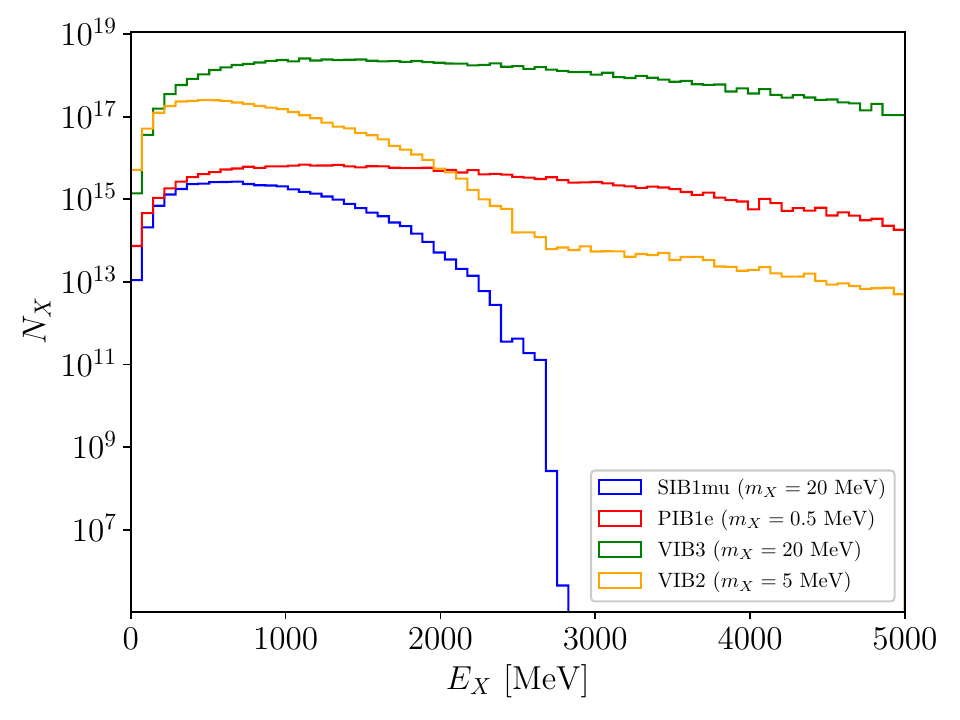}
    \caption{Energy and angular spectra for the different decay models produced from the Booster Neutrino Beam (BNB); a scalar produced through IB1 coupled to the muon (blue), a pseudoscalar produced through IB1 coupled to the electron (red), a vector produced via contact interaction IB3 (green), and a vector coupled to the pion leg via IB2 (orange). In each case we fix the couplings involved in the 3-body decay $g_\mu^S$, $g_e^P$, $g_\pi^V$ and $g_{\pi^\pm}$ to 1 and show the total number of $X=\phi, a, V$ produced from the decay.}
    \label{fig:decay_spectra}
\end{figure}

\begin{figure}[h!]
    \centering
    \includegraphics[width=0.5\textwidth]{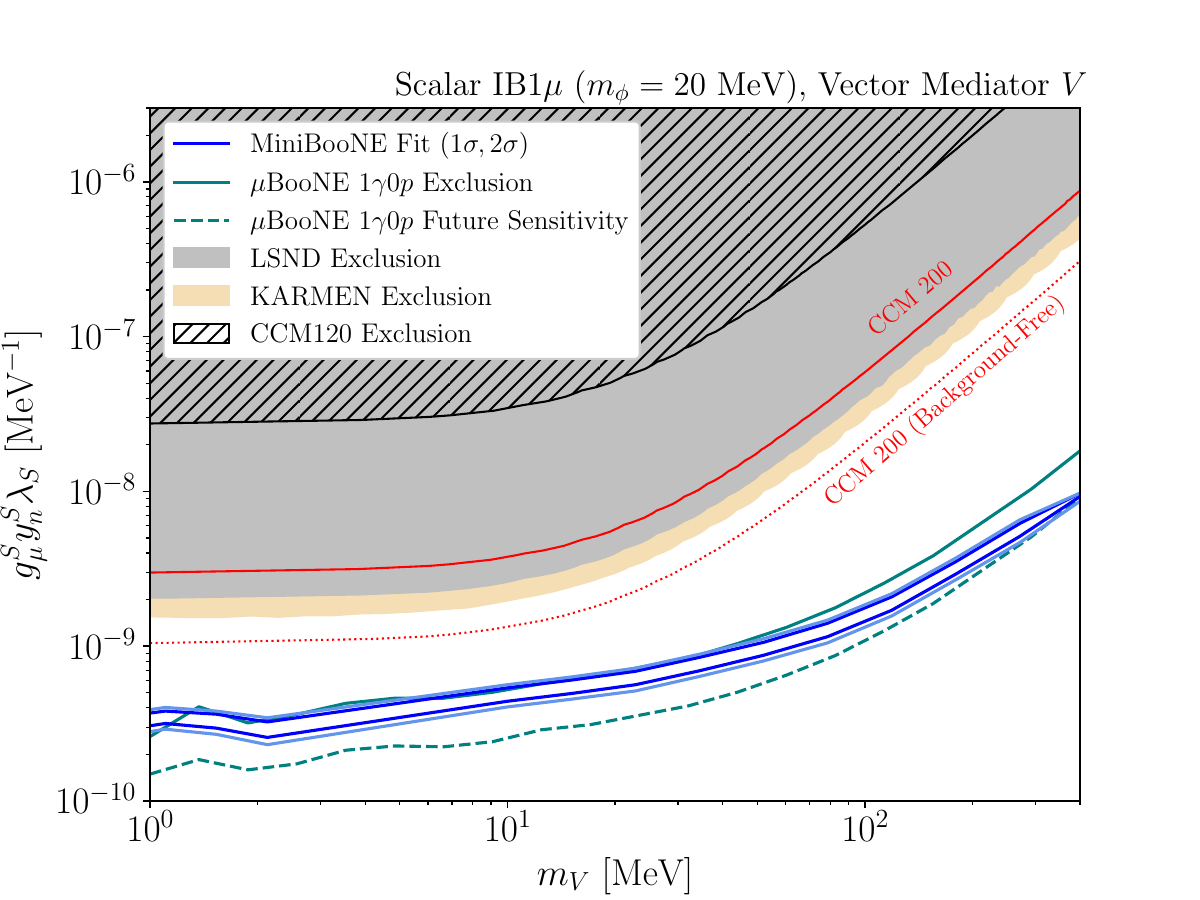}
    \caption{Parameter space for the two-mediator models, consisting of one long-lived boson produced through charged pion three-body decays into a 20 MeV scalar $\phi$ coupling to muons and scattering via a vector mediator $V$ as in Eq.~\ref{eq:IB1muS}. The exclusions (CCM120, KARMEN, LSND, and MicroBooNE) and projections for future CCM200 (in nominal background and background-free scenarios) and future MicroBooNE sensitivity are shown at 95\% C.L., while the MiniBooNE fits are shown at 68\% and 95\% C.L. in dark and light blue, respectively.}
    \label{fig:SIB1mu_add}
\end{figure}

\begin{figure*}[ht!]
    \centering
     \includegraphics[width=0.49\textwidth]{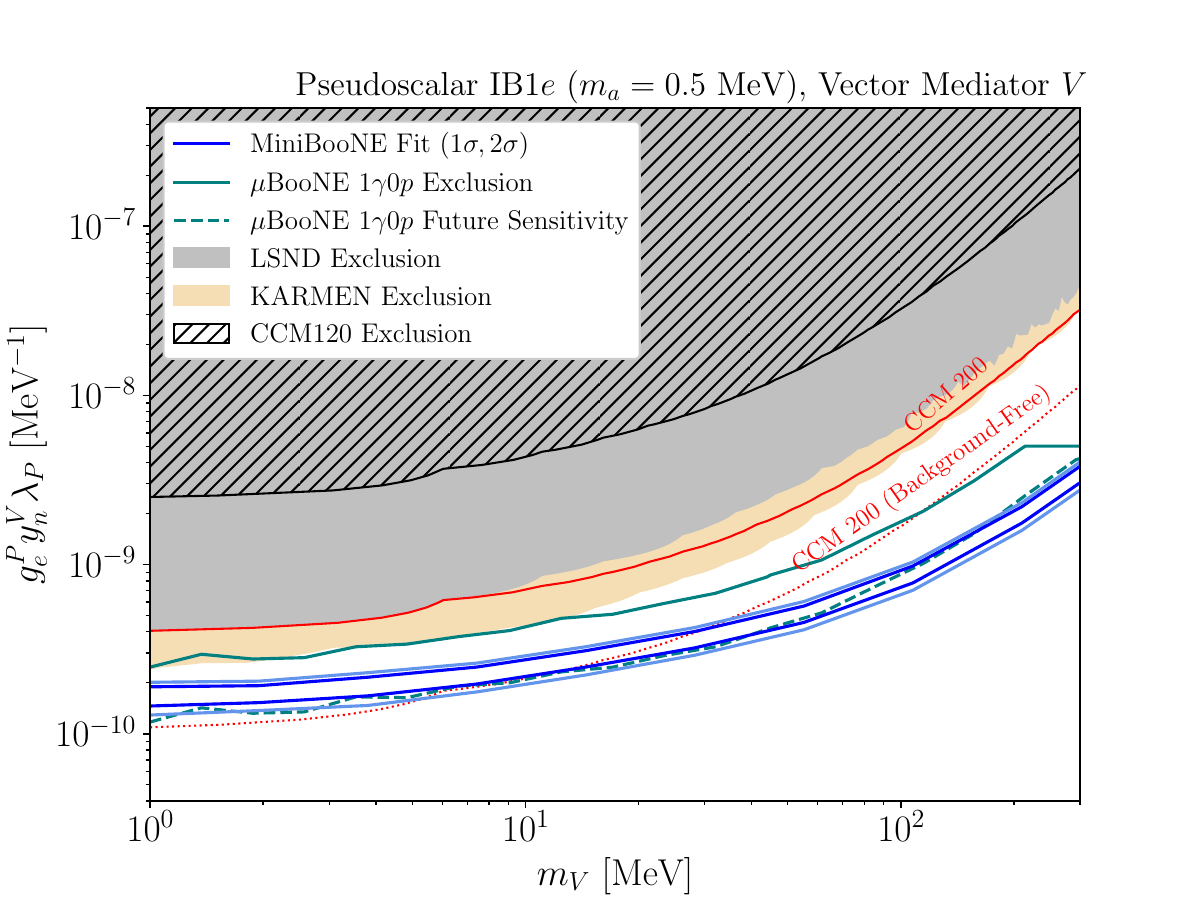}
     \includegraphics[width=0.49\textwidth]{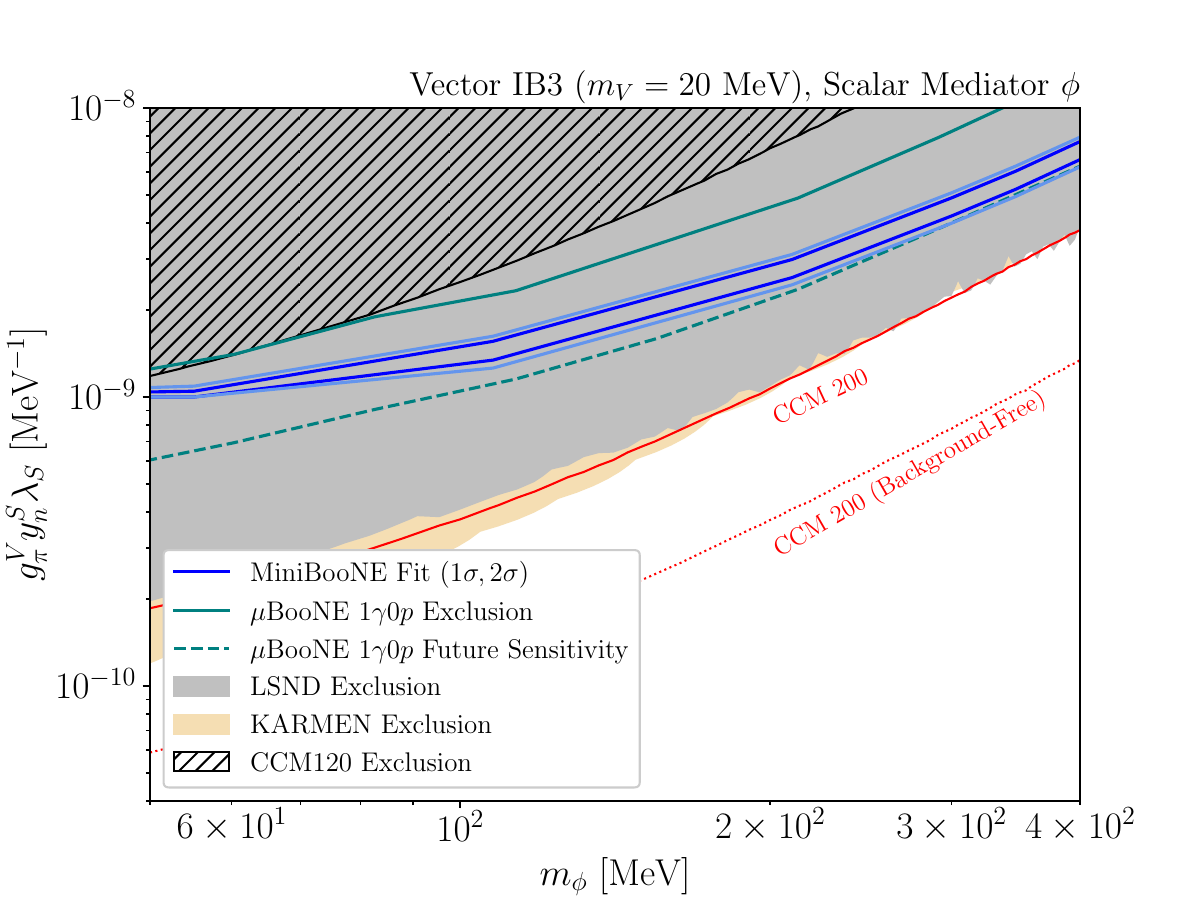}
    \caption{Same as Fig.~\ref{fig:SIB1mu_add}, but for Model II with a 0.5 MeV mass long-lived pseudoscalar $a$ scattering via a vector with mass $m_V$ as in Eq.~\ref{eq:IB1eP} (left) and Model III where a 20 MeV mass long-lived vector scatters via a scalar mediator with mass $m_\phi$ (right).}
    \label{fig:PIB1e_VIB3_add}
\end{figure*}

We first remind the reader of the phenomenological models considered in ref.~\cite{Aguilar-Arevalo:2023kvr} which involve some long-lived particle $X$, produced in charged meson decays, and a mediator particle $Y$ that facilitates the scattering into single photon final states. The 3-body decays of the mesons $\pi^\pm \to X \ell \nu$ or $K^\pm \to X \ell \nu$ (although we only consider the pion contributions in this work) can be categorized into three decay models. The models are named IB1, IB2, and IB3 (adopted from conventions in radiative meson decay literature, standing for ``internal bremsstrahlung''). We list them as follows;
\begin{itemize}
    \item IB1: radiation of $X$ off the final state charged lepton
    \item IB2: radiation of $X$ off the initial state pion
    \item IB3: radiation of $X$ from the contact interaction between the pion and leptonic currents
\end{itemize}

Since $X$ and $Y$ can take on various Lorentz representations while potentially capturing similar phenomenology and kinematic characteristics of the MiniBooNE excess, we consider a few possibilities for $X,Y$. For the representations we consider a scalar $\phi$, a pseudoscalar $a$, or a vector $V$, and take a select few low-energy effective operators to parameterize their interactions. Which representation and decay model one chooses can impact the decay kinematics, as seen by Fig.~\ref{fig:decay_spectra}. Four different configurations are itemized below and labeled as Models I-IV.

\textbf{Model I.~} Beginning with Fig.~\ref{fig:SIB1mu_add}, we consider the parameter space for a long-lived scalar particle $\phi$ produced via the IB1 decay $\pi^\pm \to \mu \nu \phi$ through a muonic coupling, and scattering through $\phi N \to \gamma N$ photoconversion via a massive vector mediator $V$. The decay and scattering are described by the phenomenological Lagrangian
\begin{equation}
\label{eq:IB1muS}
    \mathcal{L}_{int} \supset g_\mu^S \phi \bar{\mu}\mu + y_n^V V_\mu \bar{N} \gamma^\mu N - \frac{\lambda_S}{4} \phi F_{\mu\nu} H^{\mu\nu}\, + \textrm{h.c.}
\end{equation}
with $H_{\mu\nu}\equiv \partial_\mu V_\nu - \partial_\nu V_\mu$ and vector mass $m_V > m_\phi$. The event rate is proportional to the coupling product $g_\mu^S y_n^V \lambda_S$. 

In Fig.~\ref{fig:SIB1mu_add} we fix the mass of the long-lived scalar to 20 MeV and vary the mass of the mediator in the scattering $m_V$, for which we find that the fit to the MiniBooNE target and dump mode data (blue) lies around the scale  $g_\mu^S y_n^V \lambda_S \simeq 10^{-9}$ MeV$^{-1}$ at the 1$\sigma$ and 2$\sigma$ levels. The black hatched region is constrained by the CCM120 data, while constraints by LSND (primarily coming from the pion decay-in-flight analysis~\cite{LSND:1997vqj}) and KARMEN~\cite{Karmen199815} are shown in gray and tan, respectively. The constraints for LSND, KARMEN, and the CCM120 dataset are the same as those derived in ref.~\cite{Aguilar-Arevalo:2023kvr} with details therein. MicroBooNE's $1\gamma 0p$ data (solid teal) excludes more parameter space up to about a factor of 2 larger in the coupling product across all mediator masses, and we see that with a dedicated single photon search (dashed teal) and more exposure, MicroBooNE will be able to test the preferred model parameter space to explain the MiniBooNE excess. We also expect that the other SBN experiments, SBND and ICARUS, should also be very sensitive to this parameter space beyond the MiniBooNE preferred region.

\textbf{Model II.~} In Fig.~\ref{fig:PIB1e_VIB3_add} (left), the parameter space for a long-lived pseudoscalar coupling to electrons and produced through IB1 $\pi^\pm \to e \nu a$ decays is shown as a function of the coupling product and mass of a vector mediator taking place in the $a N \to \gamma N$ scattering via similar interactions,
\begin{equation}
\label{eq:IB1eP}
    \mathcal{L}_{int} \supset -ig_e^P a \bar{\mu}\gamma^5 \mu + y_n^V V_\mu \bar{N} \gamma^\mu N - \frac{\lambda_P}{4} a F_{\mu\nu} \Tilde{H}^{\mu\nu}\, + \textrm{h.c.}
\end{equation}\\

\textbf{Model III.~} In Fig.~\ref{fig:PIB1e_VIB3_add} (right) we consider a third scenario in which a massive vector mediator is long-lived and couples to the charged pion through a contact interactions, and subsequently scatters via a massive scalar mediator with mass $m_\phi > m_V$. We take the effective interaction Lagrangian
\begin{align}
\label{eq:IB3VMed}
    \mathcal{L}_{int} \supset& \, y_n^S \phi \overline{N} N - \frac{\lambda_S}{4} a F_{\mu\nu} \Tilde{H}^{\mu\nu}  \nonumber \\
    &  -ig_\pi^V \pi^+ \bar{\mu}\gamma^\rho (1-\gamma^5) \nu V_\rho \,+\, \textrm{h.c.}
\end{align}

For Models II and III, like Model I, we again see that the parameter space corresponding to the preferred region of the MiniBooNE excess should be testable with future MicroBooNE data given our assumptions here. However, in the case of Model III, we find that LSND data is already enough to exclude the explanation MiniBooNE excess under Model III beyond the 95\% C.L. region.

\textbf{Model IV.~} In the second class of phenomenological model, we consider a single long-lived vector mediator that couples to quarks and enters the pion sector via the $\chi$-PT Lagrangian in Eq.~\ref{eq:singlemedLag};
\begin{align}
\label{eq:singlemedLag}
   \mathcal{L}_{int} \supset & \, ig_{\pi^\pm} V_\mu \pi^+ (\partial^\mu \pi^-) + g_{\pi^0} \frac{e}{16\pi f_\pi}\pi^0 F_{\mu\nu}\Tilde{H}^{\mu\nu} \nonumber \\
   &- ig_{\pi NN} \pi^0 \overline{N} \gamma^5 \tau_3 N
\end{align}

\begin{figure}[h!]
    \centering
       \includegraphics[width=0.49\textwidth]{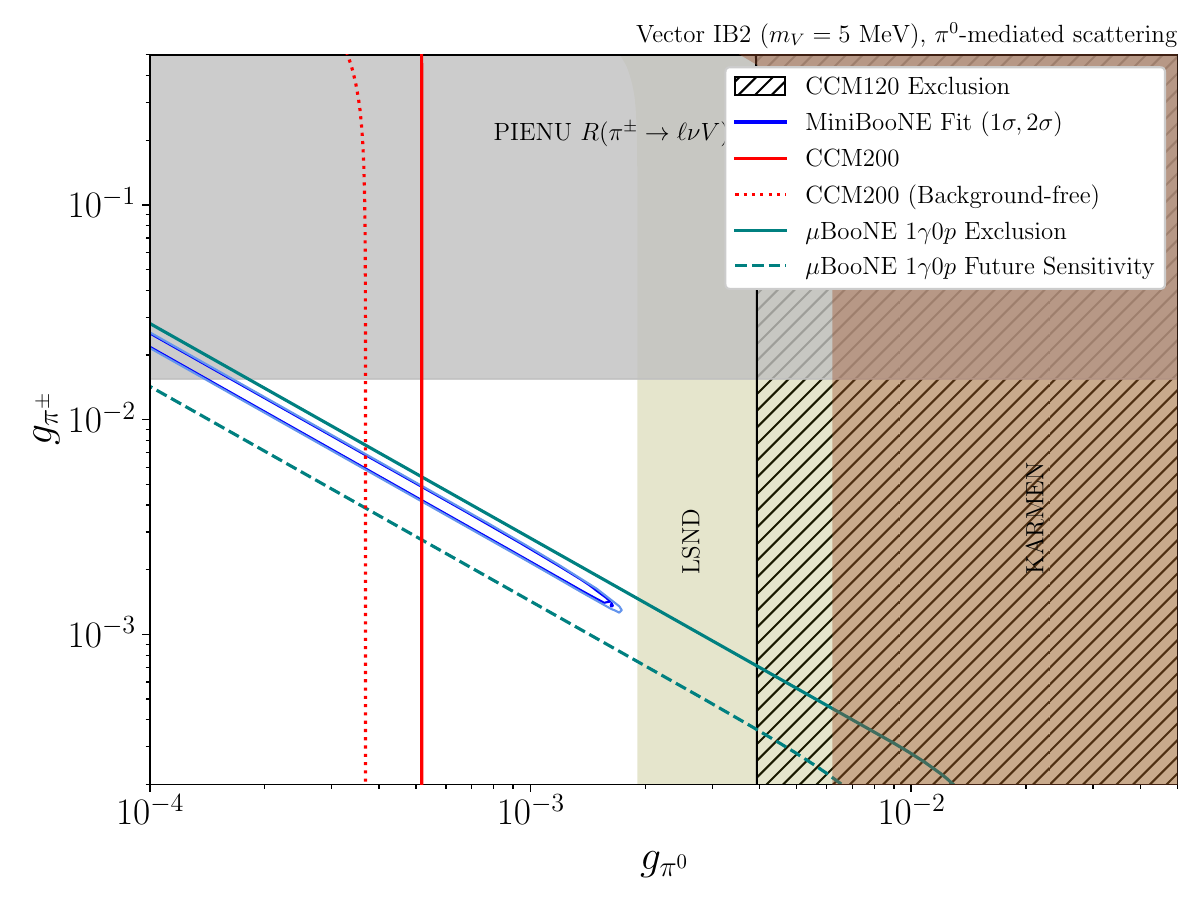}
    \caption{Parameter space for the single mediator scenario where a massive vector $V$ couples to the pion doublet via charged pion coupling $g_{\pi^\pm}$ and neutral pion coupling $g_{\pi^0}$ as in Eq.~\ref{eq:singlemedLag}. The production channels via these couplings are therefore neutral pion decay $\pi^0 \to \gamma V$ and IB2 decay $\pi^\pm \to \ell \nu V$, while the detection takes place via $\pi^0$-mediated $V N \to \gamma N$ scattering. The exclusions (CCM120, KARMEN, LSND, and MicroBooNE) and projections (CCM200 and MicroBooNE) are shown at 95\% C.L., while the MiniBooNE fits are shown at 68\% and 95\% C.L. in dark and light blue, respectively.}
    \label{fig:single_med_limits_add}
\end{figure}

\begin{figure*}[h!]
    \centering
       \includegraphics[width=0.49\textwidth]{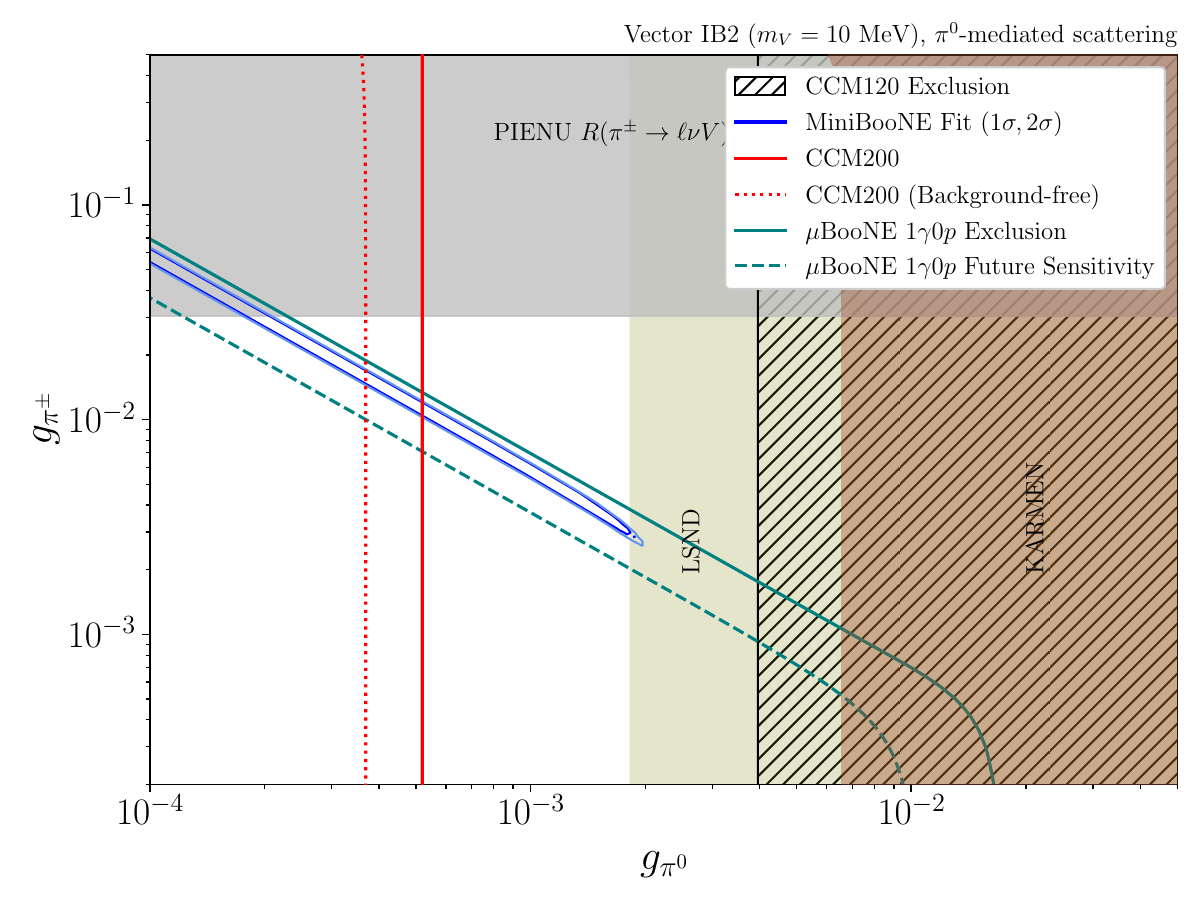}
       \includegraphics[width=0.49\textwidth]{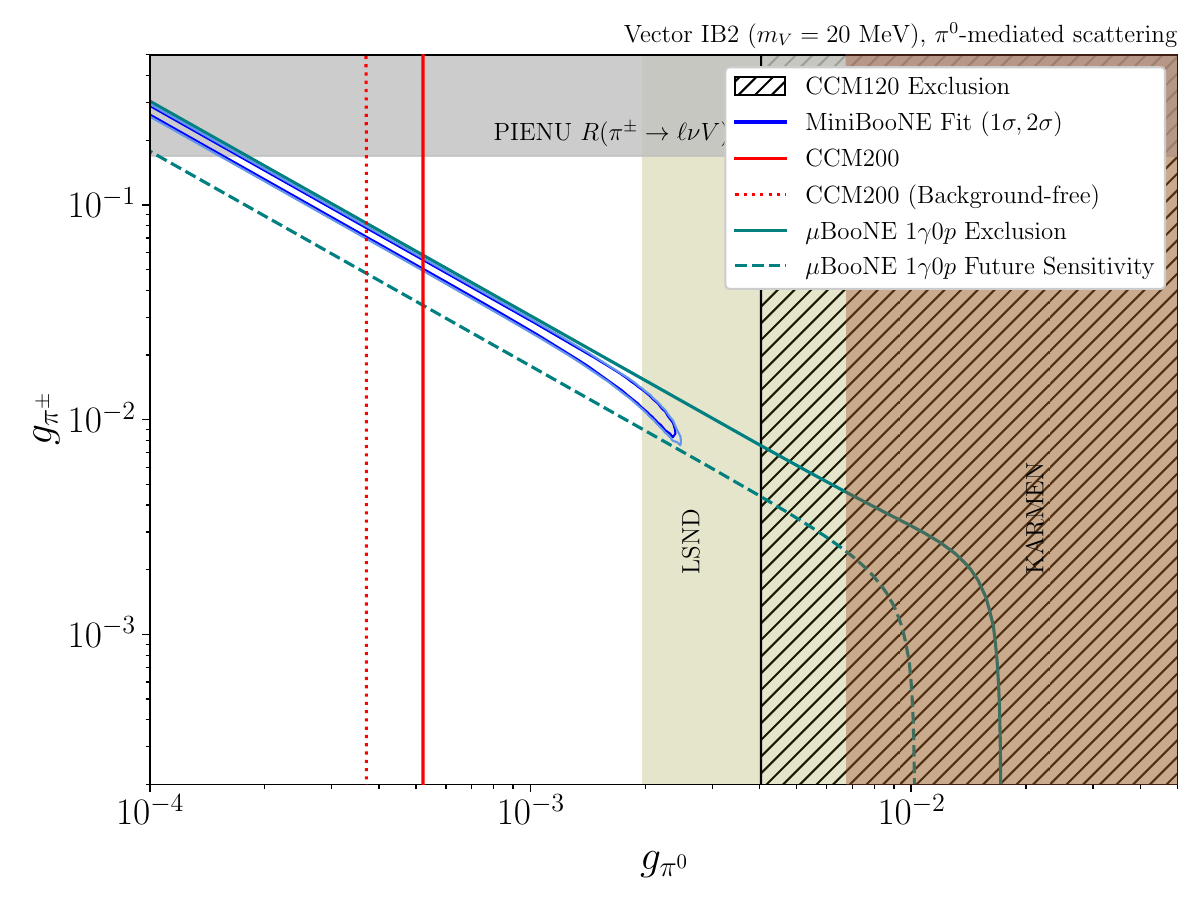}
    \caption{Same as Fig.~\ref{fig:single_med_limits_add} but for $m_V = 10$ MeV (left) and $m_V = 20$ MeV (right).}
    \label{fig:single_med_limits_10_20_add}
\end{figure*}

We show the parameter space for this model for a fixed $m_V=5,10,20$ MeV in Figs.~\ref{fig:single_med_limits_add} and ~\ref{fig:single_med_limits_10_20_add} in the $g_{\pi^\pm} - g_{\pi^0}$ plane.  The CLs for the MiniBooNE fit are shown (blue) for the combination of $\nu$, $\bar{\nu}$, and beam-dump-mode runs, exclusions set by the CCM120 engineering run are shown by the black hatched region, and future sensitivity expected in CCM200 with upgrades (red). Also shown are constraints from LSND (light yellow), KARMEN (brown), and rare charged pion decay searches from PIENU (gray). In this case we have production channels from both charged pion decays ($\pi^\pm \to \ell \nu V$), for which we take the IB2 decay mode as a benchmark, and neutral pion decays ($\pi^0 \to V \gamma$). Constraints from $\pi^\pm$ decay width measurements can be directly applied to this parameter space in terms of $g_{\pi^\pm}$, and for these we take the bounds from PIENU~\cite{PIENU:2021clt} which have set constraints on invisible radiative decays $\pi^\pm \to e^\pm \nu X$ and $\pi^\pm \to \mu \nu^\pm X$ dependent on the $X$ mass.  It is interesting to note that neutrino experiments considered here provide better limits than direct measurements, or in the case of MiniBooNE, closed contours.

This range of parameter space preferred by the MiniBooNE fit in Fig.~\ref{fig:single_med_limits} is not yet excluded by LSND at 95\% C.L., and within reach of a future $1\gamma 0 p$ search at MicroBooNE. In addition, constraints from $\pi^0$ decay width measurements apply to the $g_{\pi^0}$ coupling in this parameter space, but they apply only to larger values of the coupling that are not shown in the plot. The constraints from PIENU also begin to relax while moving to larger masses, due to the weaker branching ratio of $\pi^+\pm \to \ell \nu V$ with higher mass. In each case, we again find that a $1\gamma 0 p$ search at MicroBooNE with higher exposure should be sensitive to this explanation of the MiniBooNE excess at 95\% C.L.

\section{The MicroBooNE Prediction}
\begin{figure*}[h!]
    \centering
    \includegraphics[width=0.49\textwidth]{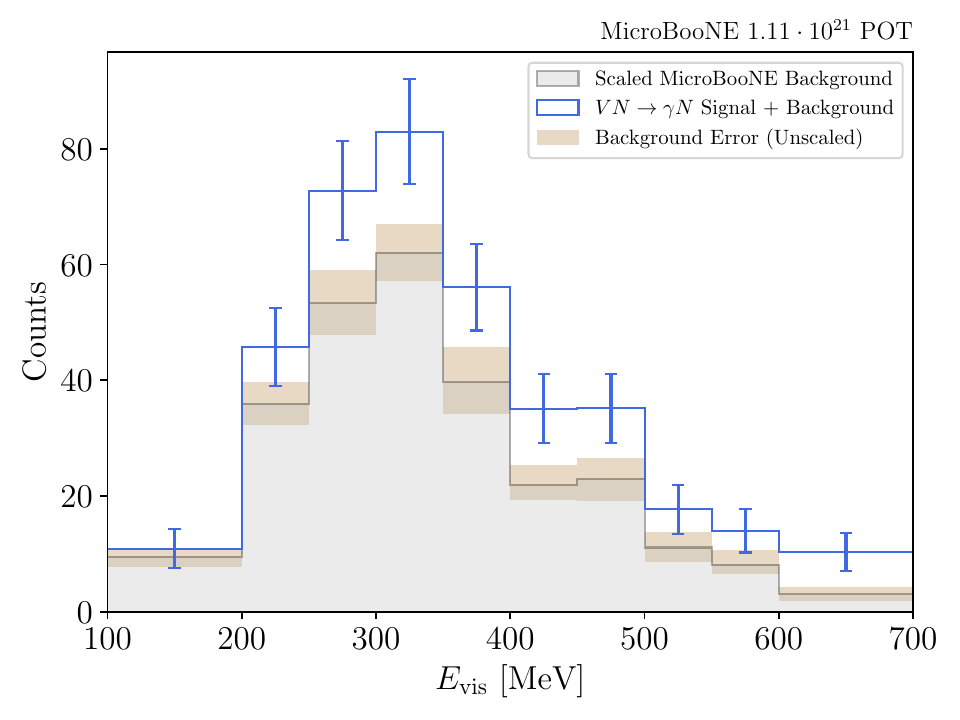}
     \includegraphics[width=0.49\textwidth]{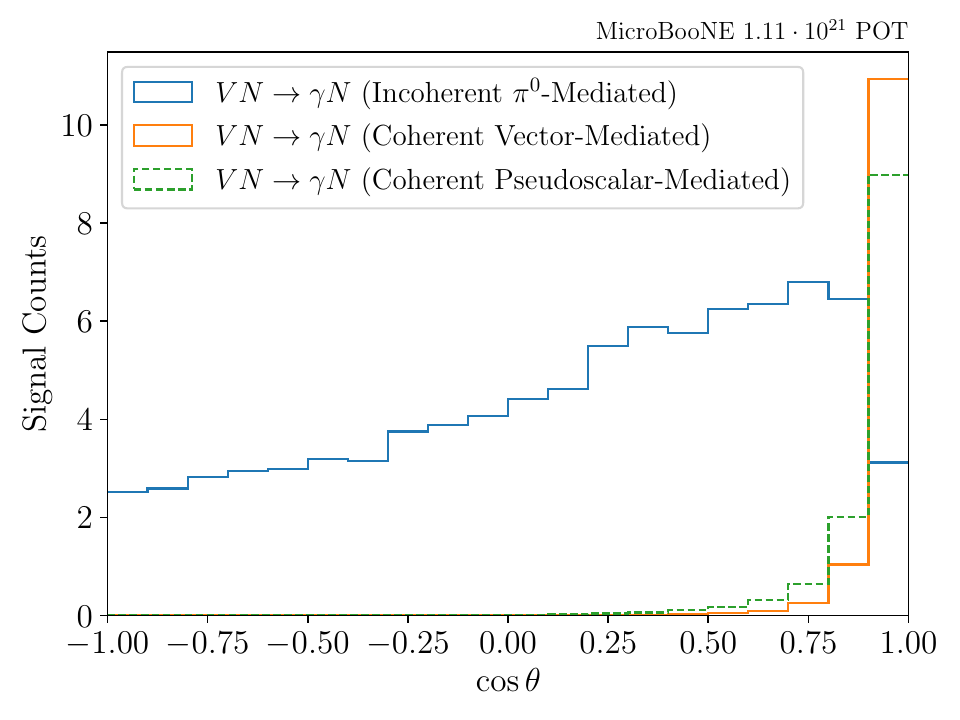}
    \caption{MicroBooNE predicted event spectra for one of the best-fit points in the model parameter space for Model IV. Left: the visible energy spectra for the signal from Primakoff-like photoconversion of the long-lived vector (blue), the scaled background prediction (gray) and scaled background error (tan) as a stacked histogram. Right: the signal prediction binned over its angular distribution in $\cos\theta$ where $\theta$ is the angle with respect to the beam axis at MicroBooNE (blue) from the same signal model shown on the left. In addition, we also show the angular distributions from coherent scattering (orange, green) that may take place in the other phenomenological models I-III.}
    \label{fig:mub_prediction}
\end{figure*}
From the scans of the parameter space in the previous section, we can forecast a possible excess that MicroBooNE could observe in the $1\gamma 0 p$ inclusive channel given the minimized $\chi^2$ of the signal hypothesis for one of the models I-IV. For example, in Model IV the predicted signal from one of the best-fit points for $m_V = 20$ MeV in the $(g_{\pi^0}, g_{\pi^\pm})$ plane is shown in Fig.~\ref{fig:mub_prediction}.

The number of contributed events from the new physics signal at MiniBooNE, MicroBooNE, and CCM are also listed in Table~\ref{tab:counts}.  It is important to note that the inverse Primakoff detection channel in liquid argon for MicroBooNE and CCM has a cross section enhancement of about three relative to MiniBooNE ($A = 12$ for carbon, $A = 40$ for argon), giving these smaller detectors improved sensitivity to this process. In Table~\ref{tab:counts} we show the excess counts and the $\chi^2$ statistic normalized to the number of degrees of freedom $\nu$ for neutrino and anti-neutrino mode in the visible energy and cosine of the angle data from MiniBooNE. For reference, the MiniBooNE visible energy and angular spectrum data (in the absence of systematic errors and covariance information) result in $\chi^2/\nu = 8.3$ ($\nu$-mode, $E_{vis}$ data), $\chi^2/\nu = 2.4$ ($\bar{\nu}$-mode, $E_{vis}$ data), $\chi^2/\nu = 9.7$ ($\nu$-mode, $\cos\theta$ data) and $\chi^2 / \nu = 1.9$ ($\bar{\nu}$-mode, $\cos\theta$ data).

In the following rows of Table~\ref{tab:counts} we also report the predicted counts at MicroBooNE with $1.11 \times 10^{21}$ POT and the $\chi^2 / \nu$ for the same point in the region preferred to fit the MiniBooNE excess. We find around 100 signal events in this model parameter space point and a similar rate in other models considered. For CCM, since the signal is dominated by the 2-body neutral pion decay, we report the event rate (without energy reconstruction smearing or signal efficiencies applied) at the $g_{\pi^0}$ point at the edge of the MiniBooNE preferred region in Fig.~\ref{fig:single_med_limits_add} for $1.5 \times 10^{22}$ POT. 

If the general class of meson portal models discussed in this paper are realized in nature, then the upcoming MicroBooNE and CCM results will complement each other and begin to paint a picture of specifically which models are in play.   This is due to the differences in the BNB and Lujan beam targets production of charged and neutral mesons and how the various models couple to them.

\begin{table*}[th!]
    \centering
    \begin{tabular}{|c|c|c|c|}
  \hline
    & Excess Counts (Theory) & $\chi^2 / \nu$ ($E_{vis})$ & $\chi^2 / \nu$  ($\cos\theta$)\\
    \hline \hline
        MiniBooNE Best-Fit ($\nu$-mode) & 456 & 3.3 & 4.2 \\ \hline
        MiniBooNE Best-Fit ($\bar{\nu}$-mode) & 134 & 1.58 & 0.81\\ \hline
        MicroBooNE 1$\gamma$0p topology, $1.11 \times 10^{21}$ POT& 113 & 3.5 & - \\ \hline
                \hline
        CCM $1.5 \times 10^{22}$ POT ($g_{\pi^0} = 1.8 \times 10^{-3}$) & 915 & - & - \\
    \hline
    \end{tabular}
    \caption{The total counts predicted by Model IV in the MiniBooNE target modes and in future MicroBooNE and CCM data given the assumptions outlined in \S~\ref{sec:mub}, along with the $\chi^2$ divided by degrees of freedom ($\nu$) for each dataset (or simulated dataset based off the null hypothesis in the case of future data). In the case of MiniBooNE data we separate the $\chi^2 / \nu$ across the visible energy ($E_{vis}$) and the angular distribution ($\cos\theta$) data. No systematic errors were assumed for the projected MicroBooNE excess or for the calculation of the reduced $\chi^2 / \nu$. Here we take an example point in model parameter space for $m_V = 5$ MeV along the band $g_{\pi^\pm} \times g_{\pi_0} \simeq 2 \times 10^{-6}$ (see Fig.~\ref{fig:single_med_limits_add}). In the last row we report the number of expected signal events at the edge of the MiniBooNE preferred contour from $\pi^0 \to V \gamma$ production at CCM200 (before applying energy reconstruction smearing and signal efficiencies).} 
    \label{tab:counts}
\end{table*}

\section{Conclusion}
In this addendum, we have extended the analysis presented in ref.~\cite{Aguilar-Arevalo:2023kvr} to include a projection for future data taken with the MicroBooNE experiment given an assumed $1.11 \times 10^{21}$ POT of collected data and $\sim 3$ times higher signal efficiency than the existing $1\gamma 0 p$ analysis already performed (ref.~\cite{MicroBooNE:2021zai}). We find that MicroBooNE is sensitive to a new physics interpretation of the MiniBooNE anomaly due to long-lived particles undergoing photoconversion in the detector via a number of possible low-energy EFTs. We estimate that if the excess is entirely due to new physics, a similar excess on the order of $\mathcal{•}{O}(100)$ events could appear in upcoming MicroBooNE data.

\section*{Acknowledgement}
We thank Aparajitha Karthikeyan for the helpful crosschecks and discussions. AAA-A, JCD and MC-E acknowledge support from DGAPA-UNAM Grant No. PAPIIT-IN104723. AT acknowledges support from the DOE
Grant No. DE-SC0010143. We acknowledge the support of the Department of Energy Office of Science, Los Alamos National Laboratory LDRD funding, and funding from the National Laboratories Office at Texas A\&M. We also wish to acknowledge support from the LANSCE Lujan Center and LANL's Accelerator Operations and Technology (AOT) division.  This research used resources provided by the Los Alamos National Laboratory Institutional Computing Program, which is supported by the U.S. Department of Energy National Nuclear Security Administration under Contract No.\,89233218CNA000001.

\bibliographystyle{apsrev4-1}
\bibliography{main}

\end{document}